\documentclass[prd,superscriptaddress,nofootinbib,amsfonts,notitlepage]{revtex4-2}
\usepackage[utf8]{inputenc}
\usepackage[T1]{fontenc}
\usepackage{hyperref}
\usepackage{graphicx,bm,amsmath,color}
\usepackage{bbold}
\usepackage{wasysym}
\usepackage{verbatim}
\usepackage{amssymb}
\usepackage{epstopdf}
\usepackage{animate}
\usepackage{xmpmulti} 
\usepackage{centernot}
\usepackage{multirow}
\usepackage{tikz}
\usepackage{graphicx}
\usepackage{float}
\usepackage{mathtools}
\usepackage{comment}
\usetikzlibrary{arrows}
\usetikzlibrary{decorations.pathreplacing}
\DeclareMathOperator{\arccoth}{arccoth}

\makeatletter

\newcommand{\be}{\begin{equation}}
\newcommand{\ee}{\end{equation}}
\newcommand{\beq}{\begin{eqnarray}}
\newcommand{\eeq}{\end{eqnarray}}
\newcommand{\ba}{\begin{align}}
\newcommand{\ea}{\end{align}}

\begin{document}

\title{Relativistic deformed   kinematics: from flat to curved spacetimes}

\author{José Javier Relancio}
\affiliation{Dipartimento di Fisica ``Ettore Pancini'', Università di Napoli Federico II, Napoli 80125, Italy;\\
INFN, Sezione di Napoli, Napoli 80125, Italy;\\
Departamento de Física Teórica and Centro de Astropartículas y Física de Altas Energías (CAPA),
Universidad de Zaragoza, Zaragoza 50009, Spain
}
\email{relancio@unizar.es}

\begin{abstract}
Doubly special relativity has been studied for the last twenty years as a way to go beyond the special relativistic kinematics, trying to capture residual effects of a quantum gravity theory. In particular, in doubly special relativity the Einstenian relativity principle is generalized, adding to the speed of light another relativistic invariant, the Planck energy. There are several papers in the literature showing a connection between this deformed kinematics and a curved momentum space. Here we review how such kinematics can be derived from geometrical ingredients in a rigorous way, and how they can be generalized when regarding a curved spacetime. For the last aim, it is mandatory to consider a particular geometry for all phase-space variables, the so-called generalized Hamilton spaces. This construction allows us to define a spacetime in these theories, which in fact depends on the momenta. Then, starting from such a momentum dependent metric, we also revise several concepts of general relativity, with the final aim of establishing a self-consistent geometrical structure from which quantum gravity phenomenology can be explored.
\end{abstract}

\maketitle

\section{Introduction}
In the history of physics the exploration of smaller scales has required new theoretical frameworks. With scientific research new theories have been developed, so more phenomena have been explained. Thanks to the technology provided by this research, smaller scales have been studied, leading to new processes that had to be understood. The problem arises when one knows that there are some missing parts or inconsistencies in the theory, while no new experimental observations are available to guide its development.
This undesirable fact emerges in theoretical physics nowadays. 

One source of inconsistencies appears when one tries to unify general relativity (GR) and quantum field theory (QFT). One of the possible issues that impedes the unification of these two theories is the role that spacetime plays in them. While in QFT spacetime is given from the very beginning, as a framework in which the processes of interactions can be described, in GR the spacetime is understood as the deformation of a flat 4-dimensional space modeled by matter and radiation. Of course, one can consider a quantum theory of gravitation where the mediation of the interaction is carried out by the graviton, a spin-2 particle, leading to Einstein’s equations~\cite{Feynman:1996kb}. The problem of this approach is that this theory is not renormalizable, and then it gives well-defined  predictions only for energies below the Planck scale.

With the huge machinery that these two theories provide we can describe, on the one hand, the massive objects (GR), and on the other one, the fundamental particles (QFT), so one could naively say that a theory containing both, a quantum gravity theory (QGT), would be completely unnecessary. But this is not the case if one wants to study the propagation and interaction of very tiny and energetic particles: the kinematics of the processes should take into account the quantum and gravitational effects together, something unthinkable if QFT and GR cannot be studied in the same framework. This kind of interactions did take place at the beginning of the universe, where a huge amount of matter was concentrated in a minute region of space. So in order to describe the first instants of the universe, a complete understanding of a QGT should be indispensable. 

Besides this, we do not know what happens inside a black hole, which is a source of contradiction between GR and QFT~\cite{Hawking:1976ra}. What happens with the information when it crosses the event horizon? If one considers that the information is lost, one is going against what quantum theory says. If on  the other hand, the information remains encrypted in the horizon surface, the evaporation of the black hole~\cite{Hawking:1974sw} would lead to a contradiction between pure and mixed states. In fact, one of the possible solutions to the information paradox, named firewall~\cite{Almheiri:2012rt} because it proposes that due to the existence of mixed states there would be particles ``burning'' an observer in free fall into the black hole, violates the equivalence principle, which states that one should not feel anything while crossing the event horizon. Another question is: what happens when one comes to the singularity? To answer all these questions, we need a QGT.

There is also a difficulty in defining spacetime. Einstein thought about being able to describe the space-time coordinates through the exchange of light signals~\cite{Einstein1905}, but when one uses this procedure, one neglects all information about the energy of the photons and assumes that the same spacetime is rebuilt by exchange of light signals of different frequencies. However, what would happen if the speed of light depends on the energy of the photon, as it happens in many theoretical frameworks which try to unify GR with QFT? In this case, the energy of the photon would affect the own structure of spacetime. Also, this procedure of identifying points of spacetime assumes that interactions are local events, happening at the same point of spacetime. This is no longer valid when one has a relativistic deformed kinematics~\cite{AmelinoCamelia:2011bm,AmelinoCamelia:2011pe}, where an energy dependent velocity and a non-locality of interactions may arise. 

Presumably, all these paradoxes and inconsistencies could be avoided if a QGT was known. Despite our ignorance about the possible consequences and implications of a complete QGT, we can pose the main properties that such a theory should have, and the characteristic phenomenological impacts that may result. 

\subsection{Noncommutative spacetime}
\label{sec:QGT}
In the last 60 years, numerous theories have tried to avoid the inconsistencies that appear when one tries to put in the same scheme GR and QFT: string theory~\cite{Mukhi:2011zz,Aharony:1999ks,Dienes:1996du}, loop quantum gravity~\cite{Sahlmann:2010zf,Dupuis:2012yw}, supergravity~\cite{VanNieuwenhuizen:1981ae,Taylor:1983su}, or causal set theory~\cite{Wallden:2010sh,Wallden:2013kka,Henson:2006kf}. The main problem is the lack of experimental observations that could tell us which is the correct theory corresponding to a QGT~\cite{LectNotes702}. 

In most of these theories, as they are trying to consider a generalization of the classical version of spacetime, a minimum length appears~\cite{Gross:1987ar,Amati:1988tn,Garay1995}, which is usually considered to be the Planck one, and then the Planck energy is taken as a characteristic energy. 

The consideration of a minimum length provokes that the spacetime acquires some very particular features that must be studied in order to know what kind of theories we are facing (see Ref.~\cite{Schiller:1996fw} for a more detailed discussion). In particular, spacetime must be quantum, discrete, and non-continuous. Due to this quantization, the concepts of a point in space and of an instant of time is lost, as a consequence of the impossibility of measuring with a greater resolution than the Planck scale. Also, this gives place to a modification of the commutation rules of space-time coordinates, because the measure of space and time leads to  non-vanishing uncertainties for position and time, $\Delta x \, \Delta t \geq l_p \, t_p$.  This  completely changes our classical notion of a Riemannian manifold, which must be substituted by an utterly concept we cannot imagine nowadays. 

 The idea of a quantum spacetime was firstly proposed by Heisenberg and Ivanenko as an attempt to avoid the ultraviolet divergences of QFT. This idea passed from Heisenberg to Peierls and to Robert Oppenheimer, and finally to Snyder, who published the first concrete example in 1947~\cite{Snyder:1946qz}. This is a Lorentz covariant model in which the commutator of two coordinates is proportional to the Lorentz generator
\begin{equation}
\left[x^\mu,x^\nu\right]\,=\,i\frac{J^{\mu \nu}}{\Lambda^2}  \,,
\end{equation}
where $\Lambda$ has dimensions of energy by dimensional arguments. But this model, originally proposed to try to rid off the ultraviolet divergences in QFT,  was forgotten when renormalization appeared as a systematic way to avoid the divergences at the level of the relations between observables. Recently, the model has been reconsidered when noncommutativity was seen as a possible approach to a QGT.  

Another widely studied model is the canonical noncommutativity~\cite{Szabo:2001kg,Douglas:2001ba}, 
\begin{equation}
\left[x^\mu,x^\nu\right]\,=\, i \Theta^{\mu \nu} \,,
\end{equation}
where $\Theta^{\mu \nu}$ is a constant matrix with dimensions of length squared. In this particular simple case of noncommutativity it has been possible to study a QFT with the standard perturbative approach.

The last model we mention here, named $\kappa$-Minkowski\footnote{We will study it in more detail in Sec.\ref{sec:DSR} and Sec.~\ref{sec:examples} as it is included in the scheme of  $\kappa$-Poincar\'{e}.}~\cite{Smolinski1994}, has the following non-vanishing commutation rules 
\begin{equation}
\left[x^0,x^i\right]\,=\,- i \frac{x^i}{\Lambda}  \,,
\end{equation}
where $\Lambda$ has also dimensions of energy.

Snyder noncommutative spacetime is very peculiar from the point of view of symmetries since the usual Lorentz transformations used in SR are still valid. But in general, in the other models of noncommutativity, linear Lorentz invariance is not a symmetry of the new spacetime, which is in agreement with what we have seen previously: the classical concept of a continuum spacetime has to be replaced somehow for Planckian scales, where new effects due to the quantum nature of gravity (for example, creation and evaporation of virtual black holes~\cite{Kallosh:1995hi}) should appear. So, while SR postulates Lorentz invariance as an exact symmetry of Nature (every experimental test up to date is in accordance with it~\cite{Kostelecky:2008ts,Long:2014swa,Kostelecky:2016pyx,Kostelecky:2016kkn}; see also the papers in~Ref.~\cite{LectNotes702}), a QGT is expected to modify  someway this symmetry. Many theories which try to describe a QGT include a modification of Lorentz invariance in one way or another (for a review, see Ref.~\cite{AmelinoCamelia:2008qg}), and the possible experimental observations that confirms or refutes this hypothesis would be very important in order to constrain these possible theories. A way to go beyond the Lorentz invariance is to consider that this symmetry would be violated for energies comparable with the high-energy scale. This is precisely what is studied in the so-called Lorentz-invariance violation (LIV) theories. In this way, the SR symmetries are only low energy approximations of the true symmetries of spacetime. We will study in the next subsection the usual theoretical framework in which these kind of theories are formulated and the main experiments where  LIV effects could be manifest.

\subsection{Lorentz Invariance Violation}

As we have previously mentioned, the symmetries of the ``classical'' spacetime could be broken at high energies due to the possible new effects of the quantum spacetime. LIV theories consider that Lorentz symmetry is violated at high energies, establishing that there is a preferred frame of reference (normally an observer aligned with the cosmic microwave background (CMB), in such a way that this radiation is isotropic). A conservative way to consider this theory is to assume the validity of the field theory framework. Then, all the terms that violate Lorentz invariance (LI) are added to the standard model (SM), leading to an effective field theory (EFT) known as the standard model extension (SME)~\cite{Colladay:1998fq} (in the simplest model one considers only operators of dimension 4 or less, known as the minimal SME, or mSME) with the condition that they do not change the field content and that the gauge symmetry is not violated.

Historically, in the middle of the past century researchers realized that LIV could have some phenomenological observations~\cite{Dirac:1951:TA,Bjorken:1963vg,Phillips:1966zzc,Pavlopoulos:1967dm,Redei:1967zz},   and in the seventies and eighties theoretical bases were settled pointing how LI could be established for low energies without being an exact symmetry at all scales~\cite{Nielsen:1978is,Ellis:1980jm,Zee:1981sy,Nielsen:1982kx,Chadha:1982qq,Nielsen:1982sz}. However, this possible way to go beyond SR did not draw much attention since it was thought that effects of new physics would only appear for energies comparable to the Planck mass. It seems impossible to talk about phenomenology of such a theory being the Planck energy of the order of $10^{19}$ GeV and having only access to energies of $10^4$  GeV from particle accelerators and  $10^{11}$ GeV from particles coming from cosmic rays. But over the past few years people have realized that there could be some effects at low energy that could find out evidences of a LIV due to amplification processes~\cite{Mattingly:2005re}. These effects were baptized as ``Windows on Quantum Gravity'' (see Refs.~\cite{Mattingly:2005re,Liberati2013} for a review).  Here we only mention some phenomenological consequences of LIV, making a distinction of one- and multi-particle sector.

There are two important cumulative effects, regarding the aforementioned amplifications mechanisms for a single particle, trying to be measured. On the one hand, if there is a deformed dispersion relation (DDR) with Lorentz invariance violating terms, the velocity of particles (and particularly photons), would depend on their energy  (this effect was considered for the first time in Ref.~\cite{Amelino-Camelia1998}). This could be measured for photons coming from a gamma-ray burst (GRB), pulsars, or active galactic nuclei (AGN), due to the long distance they travel, amplifying this possible effect. On the other hand, some terms in the mSME would produce a time delay for photons due to a helicity dependence of the velocity, phenomenon baptized as birefringence~\cite{Maccione:2008tq}.

Also, a DDR implies a modifications of the relationship between energy and momentum. This has a effect in the survey of reactions, and in particular, in their thresholds.  In this scenario, since there is a preferred reference frame, some reactions forbidden in SR are now allowed starting at some threshold energy. For example, photon splitting $\gamma \rightarrow e^+e^- $ is not allowed in usual QFT because of kinematics and charge-parity (CP) conservation, but it could be possible in a LIV scenario from some threshold energy~\cite{Jacobson:2002hd}. 

Moreover, there is as well a shift of the special relativistic thresholds. For example, the GZK cutoff~\cite{Greisen:1966jv,Zatsepin:1966jv} can be modified if a LIV scenario is present.  The GZK cutoff is a good arena for constraining LIV since the threshold of the interaction of high-energy protons with CMB photons is very sensitive to a LIV in the kinematics~\cite{Jacobson:2002hd}.

Despite the efforts of the scientific community, until now there is no clear evidences of LIV. Current experiments have only been able to put constraints in the SME parameters~\cite{Kostelecky:2008ts}. 

In this article, the main field of research is a different way to go beyond SR. In this framework there is also a high-energy scale parametrizing departures from SR, but preserving a relativity principle. 

\subsection{DSR: Doubly Special Relativity}
\label{sec:DSR}

In the previous subsection we briefly summarized the most important features of LIV. Now we can wonder if there is another option instead of violating Lorentz symmetry for going beyond SR. One could consider that Lorentz symmetry is not violated at Planckian scales but deformed. This is nothing new in physics; some symmetries have been deformed when a more complete theory, which encompasses the previous one, is considered. For example, Poincar\'{e} transformations, that are the symmetries of SR, are deformations of the Galilean transformations in classical mechanics. In this deformation, a new invariant parameter appears, the speed of light. Similarly, in a theory beyond SR (thinking on some approximation of a QGT), one could have a Poincar\'{e} deformed symmetry with a new parameter. This is what doubly special relativity (DSR) considers (see Ref.~\cite{AmelinoCamelia:2008qg} for a review).

In this theory, the Einsteinian relativity principle is generalized adding a new relativistic invariant to the speed of light $c$, the Planck energy $\Lambda$. This is why this theory is also called Doubly Special Relativity~\cite{AmelinoCamelia:2000ge,AmelinoCamelia:2000mn}.  Of course, it is assumed that in the limit in which $\Lambda$ tends to infinity, DSR becomes the standard SR.

The deformed kinematics in these theories are normally obtained from Hopf algebras~\cite{Majid:1995qg}, and a particular example is usually considered, the deformation of Poincar\'{e} symmetries through quantum algebras known as $\kappa$-Poincar\'{e}~\cite{Lukierski:1992dt,Lukierski:1993df,Majid1994,Lukierski1995}.  For the one-particle sector, one has a deformed dispersion relation and a deformed Lorentz transformation. For the two-particle system, a deformed composition law (DCL) for the momenta appears.  In $\kappa$-Poincar\'{e},  there is usually a modification of the dispersion relation and the Lorentz symmetries in the one-particle system. But this ingredient is not mandatory and there are some basis in which they are undeformed with respect to the special relativistic case~\cite{KowalskiGlikman:2002we}. Then, the main ingredient of the DSR kinematics are  a coproduct of momenta and Lorentz transformations in the two-particle system, which are non-trivial in any basis~\cite{KowalskiGlikman:2002we}. The coproduct of momenta is considered as a deformed composition law and the coproduct of the boosts tell us how the momentum of a particle changes under Lorentz transformations in presence of another particle. One of the most studied bases in $\kappa$-Poincar\'{e} is the bicrossproduct basis\footnote{For this and other bases see Refs.~\cite{KowalskiGlikman:2002we,Lukierski:1991pn}.}. The main ingredients of this basis are
\begin{equation}
\begin{split}
m^2\,&=\,\left(2 \kappa \sinh{\left(\frac{p_0}{2 \Lambda}\right)} \right)^2-\vec{p}^2 e^{p_0/\Lambda}\,,\\ 
\left[N_i,p_j\right]\,&=\, i \delta_{ij}\left(\frac{\Lambda}{2}\left(1-e^{-2p_0/\Lambda}\right)+\frac{\vec{p}^2}{2\Lambda}\right) -i \frac{p_i p_j}{\Lambda}\,, \qquad \left[N_i,p_0\right]\,=\, i p_i\,,\\ 
\Delta\left(M_i\right)\,&=\,M_i \otimes \mathbb{I}+ \mathbb{I}\otimes M_i\,, \qquad \Delta\left(N_i\right)\,=\,N_i \otimes \mathbb{I}+  e^{-p_0/\Lambda}\otimes M_i +\frac{1}{\Lambda}\epsilon_{i\,j\,k} p_j\otimes M_k\,, \\ 
\Delta\left(p_0\right)\,&=\,p_0 \otimes \mathbb{I}+ \mathbb{I}\otimes p_0\,, \qquad 
\Delta\left(p_1\right)\,=\,p_1 \otimes \mathbb{I}+ \mathbb{I}\otimes e^{-p_0/\Lambda}\,. 
\end{split}
\label{eq:coproducts}
\end{equation}

Besides the  relativistic deformed kinematics (RDK), Hopf algebras also gives the modified commutators of phase-space coordinates, and in the bicrossproduct basis of $\kappa$-Poincar\'{e}, the commutators are
\begin{equation}
\begin{split}
\left[x^0,x^i\right]\,&=\,-i\,\frac{x^i}{\Lambda} \,,\qquad
\left[x^i,x^j\right]\,=\,0\,,\qquad\left[x^0,p_0\right]\,=\,-i\,\\ 
\left[x^0,p_i\right]\,&=\,i\,\frac{p_i}{\Lambda} \,,\qquad \left[x^i,p_j\right]\,=\,-i \delta^i_j \,,\qquad \left[x^i,p_0\right]\,=\,0\,. 
\end{split}
\label{eq:pairing_intro}
\end{equation}
We see that this gives us all the ingredients that a RDK should have and also gives nontrivial commutators in phase space, making Hopf algebras an attractive way for studying DSR theories.

Regarding phenomenology, there are several (and crucial) differences from the LIV scenario. This is due to the fact that in LIV there is not an equivalence of inertial frames, which is reflected in the kinematics of these theories. In LIV, the first order correction for the threshold of a reaction is $E^3/m^2 \Lambda$~\cite{Mattingly:2005re,Liberati2013}, where $E$ is the energy of a particle involved in the process measured by our Earth-based laboratory frame, and $m$ is a mass that controls the corresponding SR threshold, so the energy has to be high enough in order to have a non-negligible correction. In contrast, in DSR there is a relativity principle, so the threshold of a reaction cannot depend on the observer; there is no new threshold for particle decays at a certain energy of the decaying particle: the energy of the initial particle is not relativistic invariant, so the threshold of such reaction cannot depend on it. Moreover, as a consequence of having a relativity principle, cancellations of effects in the deformed dispersion relation and the conservation law appear~\cite{Carmona:2010ze,Carmona:2014aba}, so many of the effects that can be observed in the LIV case  are completely invisible in this context. Also, the modification of the threshold is of the order $E/\Lambda$, implying that energies of particles involved in interactions must be comparable to the high-energy scale in order to observe an effect~\cite{Albalate:2018kcf,Carmona:2020whi,Carmona:2021lxr}. This implies that, if in these theories there is an absence of time delays  (as pointed out in~\cite{Carmona:2017oit,Carmona:2018xwm,Carmona:2019oph,Relancio:2020mpa}), which is compatible with current experimental observations~\cite{Vasileiou:2015wja,Abdalla:2019krx,Ellis:2018lca}, the high-energy scale parametrizing deviations from special relativity in DSR theories could be of the order of TeV or PeV~\cite{Albalate:2018kcf,Carmona:2020whi,Carmona:2021lxr}.

Moreover, in a collision of particles in DSR, the energy and momentum of the initial and final state particles do not fix the total initial and final momenta, since there are different channels for the reactions due to the non-symmetric  DCL (see for example~\cite{Albalate:2018kcf,Carmona:2020whi,Carmona:2021lxr}), i.e., different total momentum states would be characterized by different orderings of  momenta in the DCL. This adds an additional ingredient to take into account when studying phenomenological aspects of this theory.

Another difference with respect to LIV scenarios is that in DSR there are different basis (or representations). This is clear from the algebraic approach of Hopf algebras, where any change of momentum variables is allowed and represents the same algebra with the same properties.  Nevertheless, there is a controversy about what the physical momentum variables are~\cite{AmelinoCamelia:2010pd}. In SR we use the variables where the conservation law for momenta is the sum and where the dispersion relation is quadratic in momentum. We could wonder why no other coordinates are used. It seems a silly question in the sense that every study in SR  is easier in the usual coordinates, and the use of another (more complicated) ones would be a mess and a waste of time. But in the DSR scheme, this naive and useful argument is no longer valid. There are a lot of representations of $\kappa$-Poincaré~\cite{KowalskiGlikman:2002we}, and in some of them, the dispersion relation is the usual one, but the DCL takes a non-simple  form (the so-called ``classical basis'' is an example); however, in other bases, the DCL is a simple expression but, conversely, the DDR is not trivial (the ``bicrossproduct'' basis). So the criteria used in SR to choose the physical variables cannot be used in these schemes. From the point of view of the algebra, any basis is completely equivalent, but from the point of view of physics, it could be possible that only one should be the nature choice (supposing $\kappa$-Poincar\'{e} is the correct deformation of SR).  Ideally, one could use any momentum variable if it were possible to identify the momentum variable from a certain signal in the detector. The problem resides in the fact that the physics involved in the detection is too complicated to be able to take into account the effect of a change in momentum variables in relation with the detector signal. Maybe some physical criteria could identify the physical momentum variables. 

We have discussed in this subsection the fact that there are many ways to represent the kinematics of $\kappa$-Poincaré in different momentum variables. But in addition to this particular model, there are also a lot of them characterizing a RDK, which cannot be obtained from Hopf algebras scheme. In Sec.~\ref{sec:cmsDSR} we discuss a simple way to obtain most of these kinematics by considering a curved momentum space. The extension of these kinematics to a curved spacetime is discussed in Sec.~\ref{sec:rdkcst}. Following this line of thought, we consider in the same geometrical framework a curved momentum space and spacetime, leading to a metric in the cotangent bundle depending on all phase-space coordinates. The main ingredients and advances of this kind of geometries is posed in Sec.~\ref{sec:cbg}. Then we will address, from a geometrical perspective, a possible way to select this ``physical'' basis in Sec.~\ref{sec:einstein}, by studying the Einstein's equations in this scheme. After that, we derive the Raychaudhuri's equation and the congruence of geodesics in Sec.~\ref{sec:raychaudhuri}, so we can apply it to an expanding universe in Sec.~\eqref{sec:universe}. We see that in the basis obtained in Sec.~\ref{sec:einstein}, the second Friedmann's equation and the Raychaudhuri's one are compatible, serving us as a consistency check of our framework. We finish in Sec.~\ref{sec:conclusions} with the conclusions.

\section{Curved momentum space in DSR}
\label{sec:cmsDSR}
In the 30's, Born considered a duality between spacetime and momentum space in such a way that a curved momentum space could be also considered~\cite{Born:1938} (this idea was discussed also by Snyder some years after~\cite{Snyder:1946qz}). This proposal was postulated as an attempt to avoid the ultraviolet divergences in QFT, and until some years ago, it was not considered as a way to go beyond SR. 

In particular, in DSR scenarios there several papers showing a correspondence between a curved momentum space and RDKs. For example, in~\cite{Kowalski-Glikman:2002oyi}  a momentum geometry corresponding to de Sitter was identified by regarding the noncommutativity of $\kappa$-Minkowski. In Refs.~\cite{AmelinoCamelia:2011bm,Amelino-Camelia:2013sba,Lobo:2016blj} there are other proposals trying to establish a relation between a geometry in momentum space and a deformed kinematics. In Ref.~\cite{AmelinoCamelia:2011bm}, the DDR is defined as the squared of the distance in momentum space from the origin to a point $p$, and the DCL is associated to a non-metrical connection. In this scheme,  it is not simple to find some Lorentz transformations in the two-particle system in such a way that the relativity principle holds, the fundamental base of a RDK.

The trajectories of particles in an RDK can be depicted using a deformed Hamiltonian which encodes the dispersion relation. In this case,  a momentum dependent metric is derived from the Hamilton function~\cite{Barcaroli:2016yrl,Barcaroli:2015xda,Barcaroli:2017gvg}, a formalism known as Hamilton geometry~\cite{miron2001geometry}.  In this context, while a Lorentz transformation in the one-particle system can be defined,  the connection with a deformed addition of momenta is not clearly worked out yet.

Another proposal was presented in Ref.~\cite{Amelino-Camelia:2013sba}, achieving a different path to establish a relation between a DCL and a curved momentum space through a connection, which in this case can be  (but it is not mandatory) affine to the metric that defines the DDR in the same way as before. This link is carried out by parallel transport, implemented by a connection in momentum space, which indicates how momenta must compose. They found a way to implement some DLT implementing the relativity principle; with this procedure any connection could be considered, giving any possible RDK, and then, this would reduce to the study of a generic RDK.

In Ref.~\cite{Lobo:2016blj}, a possible correspondence between a DCL and the isometries of a curved momentum space related to translations (transformations that do not leave the origin invariant) is considered. The Lorentz transformations are the homogeneous transformations (leaving the origin invariant), in such a way that a relativity principle holds if the DDR is compatible with the DCL and this one with the DLT. As one would want 10 isometries (6 boosts and 4 translations), one should consider only maximally symmetric spaces. Then, there is only room for three options: Minkowski, de Sitter or anti-de Sitter momentum space.

However, in Ref.~\cite{Lobo:2016blj} there is not a clear way to obtain the DCL, because in fact, there are a lot of isometries that do not leave the origin invariant, so a new ingredient is mandatory. Moreover, the relativity principle argument is not really clear since one needs to talk about the transformed momenta of a set of two particles~\cite{Carmona:2012un,Carmona:2016obd}.

In the next subsection, we will see how a simple and clear identification of the kinematics can be made from the main geometrical ingredients of a curved momentum space~\cite{Carmona:2019fwf}. We present a precise way to understand a DCL: it is associated to translations, but in order to find the correct one, we must impose their generators to form a concrete subalgebra inside the algebra of isometries of the momentum space metric. 

We will see how the much studied  $\kappa$-Poincar\'{e} kinematics can be obtained from our proposal. In fact, the method we propose can be used in order to obtain other RDKs, such as Snyder~\cite{Battisti:2010sr} and the so-called hybrid models~\cite{Meljanac:2009ej}.

\subsection{Derivation of a RDK from the momentum space geometry}
\label{sec:derivation}

As we have commented previously, a RDK is composed of a DDR, a DCL and, in order to have a relativity principle, a DLT for the one- and two-particle systems, making the two previous ingredients compatible. In this section we will explain how we propose to construct a RDK from the geometry of a maximally symmetric momentum space~\cite{Carmona:2019fwf}. 

\subsubsection{Definition of the deformed kinematics}

In a maximally symmetric space there are 10 isometries. We will denote our momentum space metric as $g_{\mu\nu}(k)$. By definition, an isometry is a transformation  $k\to k'$ satisfying 
\be
  g_{\mu\nu}(k') \,=\, \frac{\partial k'_\mu}{\partial k_\rho} \frac{\partial k'_\nu}{\partial k_\sigma} g_{\rho\sigma}(k)\, .
\ee

One can always take a system of coordinates in such a way that $g_{\mu\nu}(0)=\eta_{\mu\nu}$. We write the isometries in the form
\be
k'_\mu \,=\, [T_a(k)]_\mu \,=\, T_\mu(a, k)\,, \quad\quad\quad k'_\mu \,=\, [J_\omega(k)]_\mu \,=\,J_\mu(\omega, k)\,,
\ee
where $a$ is a set of four parameters and $\omega$ of six, and 
\be
T_\mu(a, 0) \,=\, a_\mu\,, \quad\quad\quad J_\mu(\omega, 0) \,=\, 0\,,
\ee
so $J_\mu(\omega, k)$ are the 6 isometries forming a subgroup that leave the origin in momentum space invariant, and $T_\mu(a, k)$ are the other 4 isometries which transform the origin and that one can call translations.

We will identify the isometries $k'_\mu = J_\mu(\omega, k)$ with the DLT of the one-particle system, being $\omega$ the six parameters of a Lorentz transformation. The dispersion relation is defined, rather than as the square of the distance from the origin to a point $k$ (which was the approach taken in~\cite{AmelinoCamelia:2011bm}), as any arbitrary function of this distance with the SR limit when the high-energy scale tends to infinity\footnote{This disquisition can be avoided with a redefinition of the mass with the same function $f$ that relates the Casimir with the distance $ \mathcal{C}(k)=f(D(0,k))$.}. Then, under a Lorentz transformation, the equality $ \mathcal{C}(k)= \mathcal{C}(k')$ holds, allowing us to determine the Casimir directly from $J_\mu(\omega, k)$. In this way we avoid the computation of the distance and obtain in a simple way the dependence on $k$ of $ \mathcal{C}(k)$\footnote{We will see indeed a simple way to compute the distance given a metric in Sec.~\ref{subsec_action_metric}.}.

The other 4 isometries $k'_\mu = T_\mu(a, k)$ related with translations define the composition law $p\oplus q$ of two momenta $p$ and $q$, through
\be
(p\oplus q)_\mu \doteq T_\mu(p, q)\,.
\label{DCL-translations}
\ee
One can easily see that the DCL is related to the translation composition through 
\be
p\oplus q=T_p(q)=T_p(T_q(0))=(T_p \circ T_q)(0)\,.
\label{T-composition}
\ee
Note that the equation above implies that $T_{(p\oplus q)}$ differs from $(T_p \circ T_q)$ by a Lorentz transformation, since it is a transformation that leaves the origin invariant.

From this perspective, a RDK (in Sec.~\ref{sec:diagram} we will see that with this construction a relativity principle holds) can be obtained by identifying the isometries $T_a$, $J_\omega$ with the composition law and the Lorentz transformations, which fixes the dispersion relation.

Then, starting from a maximally symmetric  momentum metric, we can deduce the RDK by obtaining $T_a$, $J_\omega$ through
\be
\begin{split}
g_{\mu\nu}(T_a(k)) \,=&\, \frac{\partial T_\mu(a, k)}{\partial k_\rho} \frac{\partial T_\nu(a, k)}{\partial k_\sigma} g_{\rho\sigma}(k)\,, \\
g_{\mu\nu}(J_\omega(k)) \,=&\, \frac{\partial J_\mu(\omega, k)}{\partial k_\rho} \frac{\partial J_\nu(\omega, k)}{\partial k_\sigma} g_{\rho\sigma}(k)\,.
\end{split}
\label{T,J}
\ee
The previous equations have to be satisfied for any choice of the parameters $a$, $\omega$. From the limit $k\to 0$ in (\ref{T,J})
\be
\begin{split}
g_{\mu\nu}(a) \,=&\, \left[\lim_{k\to 0} \frac{\partial T_\mu(a, k)}{\partial k_\rho}\right] \, 
\left[\lim_{k\to 0} \frac{\partial T_\nu(a, k)}{\partial k_\sigma}\right] \,\eta_{\rho\sigma}\,,  \\
\eta_{\mu\nu} \,=&\, \left[\lim_{k\to 0} \frac{\partial J_\mu(\omega, k)}{\partial k_\rho}\right] \,  
\left[\lim_{k\to 0} \frac{\partial J_\nu(\omega, k)}{\partial k_\sigma}\right] \,\eta_{\rho\sigma}\,,
\end{split}
\ee
one can identify
\be
\lim_{k\to 0} \frac{\partial T_\mu(a, k)}{\partial k_\rho} \,=\,  \varphi_\mu^\rho(a)\,, \quad\quad\quad
\lim_{k\to 0} \frac{\partial J_\mu(\omega, k)}{\partial k_\rho} \,=\, L_\mu^\rho(\omega)\,,
\label{e,L}
\ee
where $\varphi_\mu^\alpha(k)$ is the (inverse of\footnote{Note that the metric $g_{\mu\nu}$ is the inverse of $g^{\mu\nu}$.} the) tetrad of the momentum space, and $L_\mu^\rho(\omega)$ is the standard Lorentz transformation matrix with parameters $\omega$. From Eq.~\eqref{DCL-translations} and Eq.~\eqref{e,L}, one obtains
\be
\lim_{k\to 0} \frac{\partial(a\oplus k)_\mu}{\partial k_\rho} \,=\,  \varphi_\mu^\rho(a)\,,
\label{magicformula}
\ee
which leads to a fundamental relationship between the DCL and the momentum space tetrad.

For infinitesimal transformations, we have
\be
T_\mu(\epsilon, k) = k_\mu + \epsilon_\alpha {\cal T}_\mu^\alpha(k)\,, \quad\quad\quad
J_\mu(\epsilon, k) = k_\mu + \epsilon_{\beta\gamma} {\cal J}^{\beta\gamma}_\mu(k)\,,
\label{infinit_tr}
\ee
and Eq.~(\ref{T,J}) leads to the equations
\be
\frac{\partial g_{\mu\nu}(k)}{\partial k_\rho} {\cal T}^\alpha_\rho(k) \,=\, \frac{\partial{\cal T}^\alpha_\mu(k)}{\partial k_\rho} g_{\rho\nu}(k) +
\frac{\partial{\cal T}^\alpha_\nu(k)}{\partial k_\rho} g_{\mu\rho}(k)\,,
\label{cal(T)}
\ee
\be
\frac{\partial g_{\mu\nu}(k)}{\partial k_\rho} {\cal J}^{\beta\gamma}_\rho(k) \,=\,
\frac{\partial{\cal J}^{\beta\gamma}_\mu(k)}{\partial k_\rho} g_{\rho\nu}(k) +
\frac{\partial{\cal J}^{\beta\gamma}_\nu(k)}{\partial k_\rho} g_{\mu\rho}(k)\,,
\label{cal(J)}
\ee
which allow us to obtain the Killing vectors ${\cal J}^{\beta\gamma}$, but do not completely determine ${\cal T}^\alpha$. This is due to the fact that if ${\cal T}^\alpha$, ${\cal J}^{\beta\gamma}$, are a solution of the Killing equations (\eqref{cal(T)}-\eqref{cal(J)}), then ${\cal T}^{\prime \alpha} = {\cal T}^\alpha + c^\alpha_{\beta\gamma} {\cal J}^{\beta\gamma}$ is also a solution of Eq.~(\ref{cal(T)}) for any arbitrary constants $c^\alpha_{\beta\gamma}$, and then  $T'_\mu(\epsilon, 0)=T_\mu(\epsilon, 0)=\epsilon_\mu$. This observation is completely equivalent to the comment after Eq.~\eqref{T-composition}. In order to eliminate this ambiguity, since we know that the isometry generators close an algebra, we can chose them as
\be
T^\alpha \,=\, x^\mu {\cal T}^\alpha_\mu(k), \quad\quad\quad J^{\alpha\beta} \,=\, x^\mu {\cal J}^{\alpha\beta}_\mu(k)\,,
\label{generators_withx}
\ee
so that their Poisson brackets
\begin{align}
  &\{T^\alpha, T^\beta\} \,=\, x^\rho \left(\frac{\partial{\cal T}^\alpha_\rho(k)}{\partial k_\sigma} {\cal T}^\beta_\sigma(k) - \frac{\partial{\cal T}^\beta_\rho(k)}{\partial k_\sigma} {\cal T}^\alpha_\sigma(k)\right)\,, \\
  &\{T^\alpha, J^{\beta\gamma}\} \,=\, x^\rho \left(\frac{\partial{\cal T}^\alpha_\rho(k)}{\partial k_\sigma} {\cal J}^{\beta\gamma}_\sigma(k) - \frac{\partial{\cal J}^{\beta\gamma}_\rho(k)}{\partial k_\sigma} {\cal T}^\alpha_\sigma(k)\right)\,,
\end{align}
close a particular algebra. Then, we see that this ambiguity in defining the translations is just the ambiguity in the choice of the isometry algebra, i.e., in the basis of the algebra. Every choice of the translation generators will lead to a different DCL, and then, to a different RDK.

\subsubsection{Relativistic deformed kinematics}
\label{sec:diagram}

In this subsubsection, we will prove that the kinematics obtained as proposed before is in fact a RDK. The proof can be sketched in the next diagram:
\begin{center}
\begin{tikzpicture}
\node (v1) at (-2,1) {$q$};
\node (v4) at (2,1) {$\bar q$};
\node (v2) at (-2,-1) {$p \oplus q$};
\node (v3) at (2,-1) {$(p \oplus q)^\prime$};
\draw [->] (v1) edge (v2);
\draw [->] (v4) edge (v3);
\draw [->] (v2) edge (v3);
\node at (-2.6,0) {$T_p$};
\node at (2.7,0) {$T_{p^\prime}$};
\node at (0,-1.4) {$J_\omega$};
\end{tikzpicture}
\end{center}
where the momentum with prime indicates the transformation through ${\cal J}_\omega$, and $T_p$ and $T_{p'}$, are the translations with parameters $p$ and $p'$. One can define $\bar{q}$ as the point that satisfies 
\be
(p\oplus q)' \,=\, (p' \oplus \bar{q})\,.
\label{qbar1}
\ee
One sees that in the case $q=0$, also $\bar{q}=0$, and in any other case with $q\neq 0$, the point $\bar{q}$ is obtained from $q$ by an isometry, which is a composition of  the translation $T_p$, a Lorentz transformation $J_\omega$, and the inverse of the translation $T_{p'}$ (since the isometries are a group of transformations, any composition of isometries is also an isometry). So we have found that there is an isometry  $q\rightarrow \bar{q}$, that leaves the origin invariant, and then
\be
 \mathcal{C}(q) \,=\,  \mathcal{C}(\bar{q})\,,
\label{qbar2}
\ee
since they are at the same distance from the origin. Eqs.~\eqref{qbar1}-\eqref{qbar2} imply that the deformed kinematics with ingredients $ \mathcal{C}$ and $\oplus$ is a RDK if one identifies the momenta $(p', \bar{q})$ as the two-particle Lorentz transformation of $(p, q)$. In particular, Eq.~(\ref{qbar1}) tells us that the DCL is invariant under the previously defined Lorentz transformation and Eq.~(\ref{qbar2}), together with $ \mathcal{C}(p)= \mathcal{C}(p')$, that the DDR of both momenta is also Lorentz invariant. We can see that with this definition of the two-particle Lorentz transformations, one of the particles ($p$) transforms as a single momentum, but the transformation of the other one ($q$) depends of both momenta. This computation will be carried out in the next subsection in the particular example of $\kappa$-Poincaré. 

One can notice that this framework does not allows us to construct a generic RDK where the transformations of both particles depend on both momenta. An extension to this work enabling such construction can be found in~\cite{Relancio:2021ahm}.

\subsection{Isotropic relativistic deformed kinematics}
\label{sec:examples}

In this subsection we derive the construction in detail for  $\kappa$-Poincaré kinematics. Also, we will show how to construct a RDK beyond this simple case, the kinematics known as Snyder and hybrid models. 

If the RDK is isotropic,  the general form of the algebra of the generators of isometries is  
\be
\{T^0, T^i\} \,=\, \frac{c_1}{\Lambda} T^i + \frac{c_2}{\Lambda^2} J^{0i}, \quad\quad\quad \{T^i, T^j\} \,=\, \frac{c_2}{\Lambda^2} J^{ij}\,,
\label{isoRDK}
\ee
where we assume that the generators $J^{\alpha\beta}$ satisfy the standard Lorentz algebra, and due to the fact that isometries are a group, the Poisson brackets of $T^\alpha$ and $J^{\beta\gamma}$ are fixed by Jacobi identities\footnote{The coefficients proportional to the Lorentz generators in Eq.~\eqref{isoRDK} are the same also due to Jacobi identities.}. For each choice of the coefficients $(c_1/\Lambda)$ and $(c_2/\Lambda^2)$ (and then for the algebra) and for each choice of a metric of a maximally symmetric momentum space in  isotropic coordinates, one has to obtain the isometries of such metric so that their generators close the chosen algebra in order to find a RDK. 

\subsubsection{\texorpdfstring{$\kappa$}{k}-Poincaré relativistic kinematics}
\label{subsection_kappa_desitter}
We can consider the simple case where $c_2=0$ in Eq.~\eqref{isoRDK}, so the generators of translations close a subalgebra\footnote{We have reabsorbed the coefficient $c_1$ in the scale $\Lambda$.}    
\be
\{T^0, T^i\} \,=\, \pm \frac{1}{\Lambda} T^i\,.
\label{Talgebra}
\ee
A well known result of differential geometry (see Eqs. (1.30) and (1.31) of Chapter 6 of Ref.~\cite{Chern:1999jn}) is that if the generators of left-translations $T^\alpha$, transforming $k \to T_a(k) = (a\oplus k)$, form a Lie algebra, the generators of right-translations $\tilde{T}^\alpha$, transforming $k \to (k\oplus a)$, close the same algebra but with a different sign
\be
 \{\tilde{T}^0, \tilde{T}^i\} \,=\, \mp \frac{1}{\Lambda} \tilde{T}^i \,.
\label{Ttildealgebra}
\ee   
We have found the explicit relation between the infinitesimal right-translations and the tetrad of the momentum metric in Eq.~\eqref{magicformula}, which gives
\be
(k\oplus\epsilon)_\mu\,=\,k_\mu+\epsilon_\alpha \varphi^\alpha_\mu\equiv \tilde{T}_\mu(k,\epsilon).
\ee
Comparing with Eq.~\eqref{infinit_tr} and Eq.~\eqref{generators_withx}, we see that right-translation generators are given by
\be
\tilde{T}^\alpha \,=\, x^\mu \varphi^\alpha_\mu(k)\,.
\label{Ttilde}
\ee

Since both algebras \eqref{Talgebra}-\eqref{Ttildealgebra} satisfy $\kappa$-Minkowski noncommutativity, the problem to find a tetrad $\varphi^\alpha_\mu(k)$ compatible with the algebra of Eq.~(\ref{Ttildealgebra}) is equivalent to the problem of obtaining a representation of this noncommutativity expressed in terms of canonical coordinates of the phase space. One can easily confirm that the choice of the tetrad
\be
\varphi^0_0(k) \,=\, 1\,, \quad\quad\quad \varphi^0_i(k) \,=\, \varphi^i_0(k) \,=\, 0\,, \quad\quad\quad \varphi^i_j (k) \,=\, \delta^i_j e^{\mp k_0/\Lambda}\,,
\label{bicross-tetrad}
\ee
leads to a representation of $\kappa$-Minkowski noncommutativity\footnote{See~\cite{Carmona:2021gbg,Relancio:2021ahm} for a more detailed discussion about these tetrads in momentum space and the noncommutativity of spacetime.}. 

In order to obtain the finite translations $T_\mu(a,k)$, which in this case form a group, one can try to generalize  Eq.~\eqref{e,L} to define a transformation that does not change the form of the tetrad:
\be
\varphi_\mu^\alpha(T(a, k)) \,=\, \frac{\partial T_\mu(a, k)}{\partial k_\nu} \,\varphi_\nu^\alpha(k)\,.
\label{T(a,k)}
\ee
Obviously, if $T_\mu(a,k)$ is a solution to the previous equation, it implies that the translation leaves the tetrad invariant, and then the metric, so it is therefore an isometry. Then, one can check that translations form a group since the composition of two transformations leaving the tetrad invariant also leaves the tetrad invariant. Indeed, Eq. \eqref{T(a,k)} can be explicitly solved in order to obtain the finite translations. For the particular choice of the tetrad in Eq.~\eqref{bicross-tetrad}, the translations read~\cite{Carmona:2019fwf}
\be
T_0(a, k) \,=\, a_0 + k_0, \quad\quad\quad T_i(a, k) \,=\, a_i + k_i e^{\mp a_0/\Lambda}\,,
\ee
and then the DCL is 
\be
(p\oplus q)_0 \,=\, T_0(p, q) \,=\, p_0 + q_0\,, \quad\quad\quad
(p\oplus q)_i \,=\, T_i(p, q) \,=\, p_i + q_i e^{\mp p_0/\Lambda}\,,
\label{kappa-DCL}
\ee
which is the one obtained  in the bicrossproduct basis of  $\kappa$-Poincaré kinematics~\cite{Majid1994} (up to a sign depending on the choice of the initial sign of $\Lambda$ in Eq.~\eqref{bicross-tetrad}).

From the equation
\be
\frac{\partial  \mathcal{C}(k)}{\partial k_\mu} \,{\cal J}^{\alpha\beta}_\mu(k) \,=\, 0 \,,
\label{eq:casimir_J}
\ee
one can obtain the DDR, where ${\cal J}^{\alpha\beta}$ are the infinitesimal Lorentz transformations satisfying Eq.~\eqref{cal(J)} with the metric $g_{\mu\nu}(k)=\varphi^\alpha_\mu(k)\eta_{\alpha\beta}\varphi^\beta_\nu(k)$ defined by the tetrad~\eqref{bicross-tetrad}:
\be
\begin{split}
&0 \,=\, \frac{\partial{\cal J}^{\alpha\beta}_0(k)}{\partial k_0}\,, \quad
0 \,=\, - \frac{\partial{\cal J}^{\alpha\beta}_0(k)}{\partial k_i} e^{\mp 2k_0/\Lambda} + \frac{\partial{\cal J}^{\alpha\beta}_i(k)}{\partial k_0}\,, \\
&\pm \frac{2}{\Lambda} {\cal J}^{\alpha\beta}_0(k) \delta_{ij} \,=\, - \frac{\partial{\cal J}^{\alpha\beta}_i(k)}{\partial k_j} - \frac{\partial{\cal J}^{\alpha\beta}_j(k)}{\partial k_i}\,.
\end{split}
\ee
One gets finally 
 \be
{\cal J}^{0i}_0(k) \,=\, -k_i\,, \quad \quad \quad {\cal J}^{0i}_j(k)\,=\, \pm \delta^i_j \,\frac{\Lambda}{2} \left[e^{\mp 2 k_0/\Lambda} - 1 - \frac{\vec{k}^2}{\Lambda^2}\right] \pm \,\frac{k_i k_j}{\Lambda}\,,
\label{eq:j_momentum_space}
\ee 
and then 
\be
 \mathcal{C}(k) \,=\, \Lambda^2 \left(e^{k_0/\Lambda} + e^{-k_0/\Lambda} - 2\right) - e^{\pm k_0/\Lambda} \vec{k}^2  \,,
\label{eq:casimir_momentum_space}
\ee
which is the same function of the momentum which defines the DDR of $\kappa$-Poincaré kinematics in the bicrossproduct basis~\cite{Majid1994} (up to the sign in $\Lambda$).

The last ingredient we need in order to complete the discussion of the kinematics is the two-particle Lorentz transformations. Using the diagram in Sec.~\ref{sec:diagram}, one has to find $\bar{q}$ so that
\be
(p\oplus q)' \,=\, (p'\oplus \bar{q})\,.
\ee
Equating both expressions and taking only the linear terms in $\epsilon_{\alpha\beta}$ (parameters of the infinitesimal Lorentz transformation) one arrives to the equation
\be
\epsilon_{\alpha\beta} {\cal J}^{\alpha\beta}_\mu(p\oplus q) \,=\, \epsilon_{\alpha\beta} \frac{\partial(p\oplus q)_\mu}{\partial p_\nu} {\cal J}^{\alpha\beta}_\nu(p) + \frac{\partial(p\oplus q)_\mu}{\partial q_\nu} (\bar{q}_\nu - q_\nu)\,.
\ee
From the DCL of \eqref{kappa-DCL} with the minus sign, we find
\begin{align}
& \frac{\partial(p\oplus q)_0}{\partial p_0} \,=\, 1\,, \quad
\frac{\partial(p\oplus q)_0}{\partial p_i} \,=\, 0\,, \quad
\frac{\partial(p\oplus q)_i}{\partial p_0} \,=\, - \frac{q_i}{\Lambda} e^{-p_0/\Lambda}\,, \quad
\frac{\partial(p\oplus q)_i}{\partial p_j} \,=\, \delta_i^j\,, \\
& \frac{\partial(p\oplus q)_0}{\partial q_0} \,=\, 1\,, \quad \frac{\partial(p\oplus q)_0}{\partial q_i} \,=\, 0\,, \quad
\frac{\partial(p\oplus q)_i}{\partial q_0} \,=\, 0\,, \quad \frac{\partial(p\oplus q)_i}{\partial q_j} \,=\, \delta_i^j e^{-p_0/\Lambda}\,.
\end{align}
Then, we obtain
\be
\begin{split}
\bar{q}_0 \,&=\, q_0 + \epsilon_{\alpha\beta} \left[{\cal J}^{\alpha\beta}_0(p\oplus q) - {\cal J}^{\alpha\beta}_0(p)\right]\,, \\
\bar{q}_i \,&=\, q_i + \epsilon_{\alpha\beta} \, e^{p_0/\Lambda} \, \left[{\cal J}^{\alpha\beta}_i(p\oplus q) - {\cal J}^{\alpha\beta}_i(p) + \frac{q_i}{\Lambda} e^{-p_0/\Lambda} {\cal J}^{\alpha\beta}_0(p)\right]\,,
\end{split}
\label{eq:jr_momentum_space}
\ee
and one can check that this is the Lorentz transformation of the two-particle system of $\kappa$-Poincaré in the bicrossproduct basis~\eqref{eq:coproducts}~\cite{Majid1994}.

For the choice of the tetrad in Eq.~\eqref{bicross-tetrad}, the metric in momentum space reads~\footnote{This is the de Sitter metric written in the comoving coordinate system used in Refs.~\cite{Carmona:2019fwf,Gubitosi:2013rna}.}
\be
g_{00}(k) \,=\, 1\,, \quad\quad\quad g_{0i}(k) \,=\, g_{i0}(k) \,=\, 0\,, \quad\quad\quad g_{ij}(k) \,=\, - \delta_{ij} e^{\mp 2k_0/\Lambda}\,.
\label{bicross-metric}
\ee
Computing the Riemann-Christoffel tensor, one can check that it corresponds to a de Sitter momentum space with curvature $(12/\Lambda^2)$. In~\cite{Carmona:2019fwf} it was shown that the way we have constructed the RDK as imposing the invariance of the tetrad cannot be followed for the case of anti-de Sitter space. 

To summarize, we have found the $\kappa$-Poincaré kinematics in the bicrossproduct basis~\cite{KowalskiGlikman:2002we} from geometric ingredients of a de Sitter momentum space with the choice of the tetrad of Eq.~\eqref{bicross-tetrad}. For different choices of tetrad (in such a way that the generators of Eq.~\eqref{Ttilde} close the algebra Eq.~\eqref{Ttildealgebra}), one will find the $\kappa$-Poincaré kinematics in different bases. Then, the different bases of the deformed kinematics are just different choices of coordinates in de Sitter space. Note that when generators of right-translations constructed from the momentum space tetrad close the algebra of Eq.~\eqref{Ttildealgebra}, the DCL obtained is associative (this can be easily understood since as the generators of translations close an algebra~\eqref{Talgebra}, translations form a group).

\subsubsection{Beyond \texorpdfstring{$\kappa$}{k}-Poincaré relativistic kinematics}

The other simple choice in the algebra of the translation generators is $c_1=0$, leading to the Snyder algebra explained in the introduction. As discussed in~\cite{Carmona:2019fwf}, from the isometries associated to translations one is able to find the DCL of Snyder kinematics~\cite{Battisti:2010sr} (the first order terms were obtained also in Ref.~\cite{Banburski:2013jfa}).

When both coefficients, $c_1$ and $c_2$, are non-zero, one has the algebras of the generators of translations  known as hybrid models~\cite{Meljanac:2009ej}. The DCL in these cases can be obtained from a power expansion in $(1/\Lambda)$ asking to be an isometry and that their generators close the desired algebra. With this procedure, one can get the same kinematics found in Ref.~\cite{Meljanac:2009ej}.  

The DCL obtained when the generators of translations  close a subalgebra  (the case of $\kappa$-Poincaré) is the only one which is associative. The other compositions obtained when the algebra is Snyder or any hybrid model do not have this property.  This is an important difference between the algebraic and geometric approaches: the only isotropic RDK obtained from the Hopf algebra approach is $\kappa$-Poincaré, since one asks  the generators of translations to close an algebra (and then, one finds an associative composition of momenta), eliminating any other option. With this proposal, identifying a correspondence between translations of a maximally symmetric momentum space whose generators close a certain algebra and a DCL, we open up the possibility to construct more RDKs in a simple way. 

\subsection{A comment about multi-particle states and relative locality}

As commented previously, the main ingredient of RDKs is a DCL, and therefore, it is mandatory to consider a multi-particle system. Our prescription of constructing a curved momentum space could be seen as a simple way to construct RDKs by geometrical arguments in which we did not face neither how spacetime arises in this scheme for a single particle (which will be discussed in the following), nor how to describe the spacetime of several particles involved in an interaction (as originally proposed in the relative locality framework~\cite{AmelinoCamelia:2011bm}). Nevertheless, in~\cite{Relancio:2021ahm} we proposed a possible way to consider the geometric setup for several particles by relying in the phase-space construction of Sec.~\ref{sec:cbg}. 

Indeed, the relative locality principle can be derived in this geometrical context, describing the position of the particles before and after a collision as a function of the interaction vertex  and the momenta involved in the process. This construction is fully developed for the flat space-time scenario and for more than two particles, but its generalization to a curved regime is still an open problem. 

\section{Relativistic deformed kinematics in curved spacetimes}
\label{sec:rdkcst}
In this section we discuss how to consider a RDK in curved spacetimes~\cite{Relancio:2020zok,Pfeifer:2021tas}. While this could be seen as a purely curiosity-driven study, there are several scenarios for which such structure is mandatory. The description of the first instants of the Universe involve high-energy particles in a strong gravitational field, so such kinematics could change these interactions. Also, it is necessary for identifying a possible time delay with a consistent description of the propagation of massless particles from the astrophysical sources to our telescopes which takes into account a deformed kinematics in an expanding Universe   \cite{Barcaroli:2016yrl,Jacob:2008bw,Rosati:2015pga,Amelino-Camelia:2016ohi,Pfeifer:2018pty}. Moreover, as pointed out in~\cite{Pfeifer:2021tas}, these RDKs can play a crucial role in astrophysical setups, such as the collisional Penrose processes~\cite{Banados:2009pr,Grib:2010dz,Hackmann:2020ogy,Liberati:2021uom}: two particles colliding near the horizon of a black hole could extract energy from the black hole much more efficiently that the originally process suggested by Penrose~\cite{Penrose:1971uk}. For this particular survey, a well-defined notion of RDK in curved spacetimes is necessary, and it is expected that the result obtained in GR will be modified. 

While there are several papers in the literature addressing this issue~\cite{Barcaroli:2016yrl,Barcaroli:2015xda,Barcaroli:2017gvg,Letizia:2016lew}, they only faced the DDR, without considering the possible modification on the DCL arising when considering a curved spacetime.

\subsection{Construction of the metric in phase space}
\label{sec:construction_metric_ps}
Here we explore a simple way to generalize the results obtained in the previous section by taking into account a curvature of spacetime, characterized by a metric $g_{\mu\nu}^x(x)$. For that aim, we start with the action of a free particle in SR 
\begin{equation}
S\,=\,\int{\left(\dot{x}^\mu k_\mu-\mathcal{N} \left( \mathcal{C}(k)-m^2\right)\right)d\tau}\,,
\label{eq:SR_action}
\end{equation}
where $ \mathcal{C}(k)=k_\alpha \eta^{\alpha\beta }k_\beta$, $\mathcal{N}$ is a Lagrange multiplier imposing the mass-shell condition, and $\tau$ plays the role of the proper time or the affine parametrization depending if one is considering a massive or a
massless particle, respectively. It is easy to check that the same equation of motion derived  in GR by solving the geodesic equation can be obtained just by replacing $\bar{k}_\alpha=\bar{e}^\nu_\alpha (x) k_\nu$\footnote{To avoid confusions, we will use the symbol $\bar{e}$ in order to denote the inverse of the tetrad.} in Eq.~\eqref{eq:SR_action}
\begin{equation}
S\,=\,\int{\left(\dot{x}^\mu k_\mu-\mathcal{N} \left( \mathcal{C}(\bar{k})-m^2\right)\right)d\tau}\,,
\label{eq:GR_action}
\end{equation}
where $\bar{e}^\nu_\alpha(x)$ is the inverse of the tetrad of the space-time metric  $e_\nu^\alpha(x)$, so that 
\begin{equation}
g^x_{\mu\nu}(x)\,= \, e^\alpha_\mu (x) \eta_{\alpha\beta} e^\beta_\nu (x)\,,
\label{eq:metric-st}
\end{equation}
and then, the dispersion relation is 
\begin{equation}
 \mathcal{C}(\bar{k})\,=\,\bar{k}_\alpha \eta^{\alpha\beta }\bar{k}_\beta\,=\,k_\mu g_x^{\mu\nu}(x) k_\nu\,.
\label{eq:cass_GR}
\end{equation}

As we saw in the previous section, the dispersion relation can be interpreted as a function of the distance in momentum space from the origin to a point $k$, so one can consider the following line element for momenta 
\begin{equation}
d\sigma^2\,=\,dk_{ \alpha}g_k^{\alpha\beta}(k)dk_{ \beta}\,=\,dk_{ \alpha}\bar{\varphi}^\alpha_\gamma(k)\eta^{\gamma\delta}\bar{\varphi}^\beta_\delta(k)dk_{ \beta}\,,
\label{eq:line_m1}
\end{equation}
where we use the same notation as before, being $\bar{\varphi}^\alpha_\beta(k)$  the inverse of the tetrad in momentum space, $\varphi^\alpha_\beta(k)$. This can be easily extended to the curved space-time case introducing the variables $\bar{k}$ in the previous momentum line element, obtaining
\begin{equation}
d\sigma^2\,\coloneqq\,d \bar{k}_{\alpha}g_{\bar{k}}^{\alpha\beta}( \bar{k})d \bar{k}_{ \beta}\,=\,dk_{\mu}g^{\mu\nu}(x,k)dk_{\nu}\,,
\end{equation}
where in the second equality we have taken into account that the distance is computed along a fiber, i.e., the Casimir is viewed as some function of the squared distance from the point $(x,0)$ to $(x,k)$ (we will see this in more detail in Secs.~\ref{subsec_action_metric} and \ref{sec:casimir_horizontal}). The metric tensor $g^{\mu\nu}(x,k)$ in momentum space depending on space-time coordinates is constructed with the tetrad of spacetime and the original metric in momentum space, explicitly
\begin{equation}
g_{\mu\nu}(x,k)\,=\,\Phi^\alpha_\mu(x,k) \eta_{\alpha\beta}\Phi^\beta_\nu(x,k)\,,
\label{eq:cotangent_metric_tetrads}
\end{equation}
where 
\begin{equation}
\Phi^\alpha_\mu(x,k)\,=\,e^\lambda_\mu(x)\varphi^\alpha_\lambda(\bar{k})\,.
\end{equation}
As pointed out in~\cite{Relancio:2020zok}, in the way we have constructed this metric, it is invariant under space-time diffeomorphisms. The momentum changes of coordinates will be discussed in Sec.~\ref{sec:choice_mb}. 
\subsection{Momentum isometries in curved spacetime}

Now we will see that, from the definition of the metric, one can easily generalize for a curved spacetime the momentum transformations obtained in the previous section for a maximally symmetric momentum space: as in the flat space-time case, there are still 10 momentum isometries for a fixed spacetime point $x$, 4  translations and 6 transformations leaving the point in phase space $(x,0)$ invariant, and we can also understand the dispersion relation as a function of the squared distance from the point $(x,0)$ to $(x,k)$. 

\subsubsection{Modified  translations}

In the previous section we have found the translations from Eq.~\eqref{T(a,k)}
\begin{equation}
\varphi^\mu_\nu(p\oplus q) \,=\, \frac{\partial (p\oplus q)_\nu}{\partial q_\rho} \, \varphi^\mu_\rho(q)\,,
\end{equation}
so we should find the new translations by replacing $p\rightarrow \bar{p}_\mu=\bar{e}_\mu^\nu(x) p_\nu$, $q\rightarrow \bar{q}_\mu=\bar{e}_\mu^\nu(x) q_\nu$ on it
\begin{equation}
\varphi^\mu_\nu(\bar{p} \oplus \bar{q}) \,=\,  \frac{\partial (\bar{p} \oplus \bar{q})_\nu}{\partial \bar{q}_\rho} \, \varphi_\rho^{\,\mu}( \bar{q})\,.
\label{eq:tetrad_composition2}
\end{equation}
This leads us to introduce a generalized composition law ($\bar{\oplus}$) for a curved spacetime such that 
\be
(\bar{p} \oplus \bar{q})_\mu \,=\, \bar{e}_\mu^\nu(x) (p \bar{\oplus} q)_\nu\,.
\label{eq:composition_cotangent}
\ee
Then, one has 

\begin{equation}
\begin{split}
e^\tau_\nu(x)\varphi^\mu_\tau(\bar{p} \oplus \bar{q}) \,=&\,e^\tau_\nu(x)\frac{\partial (\bar{p} \oplus \bar{q})_\tau}{\partial \bar{q}_\sigma}\varphi_\sigma^{\,\mu}(\bar{q})\,=\,e^\tau_\nu(x) \bar{e}^\lambda_\tau(x) \,\frac{\partial (p \bar{\oplus} q)_\lambda}{\partial \bar{q}_\sigma}\varphi_\sigma^{\,\mu}(\bar{q})\\
=&\,\frac{\partial (p \bar{\oplus} q)_\nu}{\partial\bar{q}_\sigma} \varphi_\sigma^{\,\mu}(\bar{q})\,=\, 
\frac{\partial (p \bar{\oplus} q)_\nu}{\partial q_\rho}\frac{\partial q_\rho}{\partial\bar{q}_\sigma} \varphi_\sigma^{\,\mu}(\bar{q})\,=\,
\frac{\partial (p \bar{\oplus} q)_\nu}{\partial q_\rho} e_\rho^\sigma(x) 
\varphi_\sigma^{\,\mu}(\bar{q})\,,
\end{split}
\end{equation}
i.e.,
\begin{equation}
\Phi^\mu_\nu(x,(p \bar{\oplus} q)) \,=\,  \frac{\partial (p \bar{\oplus} q)_\nu}{\partial q_\rho} \, \Phi_\rho^{\,\mu}(x,q)\,.
\label{eq:tetrad_cotangent}
\end{equation}
We have obtained, for a fixed $x$, the momentum isometries of the metric leaving the form of the tetrad invariant in the same way we did in the previous section. 

As we saw in the same section, the translations defined in this way form by construction a group, so the composition law must be associative. We can now show that the barred composition law is also associative from the fact that  the composition law $\oplus$ is associative. If we define $\bar{r}=(\bar{k}\oplus \bar{q})$ and $\bar{l}=(\bar{p}\oplus \bar{k})$, then we have  $r=(k \bar{\oplus} q)$ and $l=(p \bar{\oplus} k)$. Hence
\begin{equation}
(\bar{p}\oplus \bar{r})_\mu\,=\,\bar{e}^\alpha_\mu(x)(p\bar{\oplus}r)_\alpha\,=\,\bar{e}^\alpha_\mu(x)(p\bar{\oplus}(k\bar{\oplus}q))_\alpha\,,
\end{equation}
and 
\begin{equation}
(\bar{l}\oplus \bar{q})_\mu\,=\,\bar{e}^\alpha_\mu(x)(l\bar{\oplus}q)_\alpha\,=\,\bar{e}^\alpha_\mu(x)((p\bar{\oplus}k)\bar{\oplus}q)_\alpha\,,
\end{equation}
but due to the associativity of $\oplus$
\begin{equation}
(\bar{p}\oplus \bar{r})_\mu\,=\,(\bar{l}\oplus \bar{q})_\mu\,,
\end{equation}
and then 
\begin{equation}
(p\bar{\oplus}(k\bar{\oplus}q))_\alpha\,=\,((p\bar{\oplus}k)\bar{\oplus}q)_\alpha\,,
\end{equation}
we conclude that $\bar{\oplus}$ is also associative. We have then shown that in the cotangent bundle with a constant scalar of curvature in momentum space, one can also define associative momentum translations. 

\subsubsection{Modified Lorentz transformations}
One can also replace $k$ by $\bar{k}_\mu=\bar{e}_\mu^\nu(x) k_\nu$ in  Eq.~\eqref{cal(J)} 
\be
\frac{\partial g^k_{\mu\nu}(k)}{\partial k_\rho} {\cal J}^{\beta\gamma}_\rho(k) \,=\,
\frac{\partial{\cal J}^{\beta\gamma}_\mu(k)}{\partial k_\rho} g^k_{\rho\nu}(k) +
\frac{\partial{\cal J}^{\beta\gamma}_\nu(k)}{\partial k_\rho} g^k_{\mu\rho}(k)\,,
\ee
obtaining 
\be
\frac{\partial g^{\bar{k}}_{\mu\nu}(\bar{k})}{\partial \bar{k}_\rho} {\cal J}^{\beta\gamma}_\rho(\bar{k}) \,=\,
\frac{\partial{\cal J}^{\beta\gamma}_\mu(\bar{k})}{\partial \bar{k}_\rho} g^{\bar{k}}_{\rho\nu}(\bar{k}) +
\frac{\partial{\cal J}^{\beta\gamma}_\nu(\bar{k})}{\partial \bar{k}_\rho} g^{\bar{k}}_{\mu\rho}(\bar{k})\,.
\ee
From here, we have
\begin{equation}
\frac{\partial g^{\bar{k}}_{\mu\nu}(\bar{k})}{\partial k_\sigma}e^\rho_\sigma(x) {\cal J}^{\alpha\beta}_\rho(\bar{k}) \,=\,
\frac{\partial{\cal J}^{\alpha\beta}_\mu(\bar{k})}{\partial k_\sigma}e^\rho_\sigma(x) g^{\bar{k}}_{\rho\nu}(\bar{k}) +
\frac{\partial{\cal J}^{\alpha\beta}_\nu(\bar{k})}{\partial k_\sigma}e^\rho_\sigma(x) g^{\bar{k}}_{\mu\rho}(\bar{k})\,.
\end{equation}
Multiplying by $e_\lambda^\mu(x)e_\tau^\nu(x)$ both sides of the previous equation one finds 
\begin{equation}
\frac{\partial g_{\lambda \tau}(x,k)}{\partial k_\rho} \bar{{\cal J}}^{\alpha\beta}_\rho(x,k) \,=\,
\frac{\partial\bar{{\cal J}}^{\alpha\beta}_\lambda(x,k)}{\partial k_\rho} g_{\rho\tau}(x,k) +
\frac{\partial\bar{{\cal J}}^{\alpha\beta}_\tau (x,k)}{\partial k_\rho}g_{\lambda\rho}(x,k)\,,
\end{equation}
where we have defined
\begin{equation}
 \bar{{\cal J}}^{\alpha\beta}_\mu(x,k) \,=\,e^\mu_\nu(x){\cal J}^{\alpha\beta}_\nu(\bar{k})\,.
\end{equation}
We see that $ \bar{{\cal J}}^{\alpha\beta}_\mu(x,k)$ are the new isometries of the momentum metric that leave the momentum origin invariant for a fixed point $x$.
\subsubsection{Modified dispersion relation}

With our prescription, the generalization to Eq.~\eqref{eq:casimir_J} 
\be
\frac{\partial  \mathcal{C}(k)}{\partial k_\mu} \,{\cal J}^{\alpha\beta}_\mu(k) \,=\, 0\, ,
\ee
in presence of a curved spacetime is 
 \begin{equation}
\frac{\partial  \mathcal{C}(\bar{k})}{\partial \bar{k}_\mu}{\cal J}^{\alpha\beta}_\mu(\bar{k})\,=\,0\,.
\end{equation}
The generalized infinitesimal Lorentz transformation in curved spacetime, defined by $\bar{{\cal J}}^{\alpha\beta}_\lambda(x,k)$, when acting on $ \mathcal{C}(\bar{k})$ is
\be
\begin{split}
\delta  \mathcal{C}(\bar{k}) \,=&\, \omega_{\alpha\beta} \frac{\partial  \mathcal{C}(\bar{k})}{\partial k_\lambda}\,\bar{{\cal J}}^{\alpha\beta}_\lambda(x,k) \,=\, \omega_{\alpha\beta} \frac{\partial  \mathcal{C}(\bar{k})}{\partial \bar{k}_\rho}\,\frac{\partial\bar{k}_\rho}{\partial k_\lambda}\,\bar{{\cal J}}^{\alpha\beta}_\lambda(x,k) \\
=&\,\omega_{\alpha\beta} \frac{\partial  \mathcal{C}(\bar{k})}{\partial \bar{k}_\rho}\,\bar{e}^\lambda_\rho(x)\,\bar{{\cal J}}^{\alpha\beta}_\lambda(x,k) \,=\, 
\omega_{\alpha\beta} \frac{\partial  \mathcal{C}(\bar{k})}{\partial \bar{k}_\rho}\,{\cal J}^{\alpha\beta}_\rho(\bar{k}) \,=\, 0.
\end{split}
\ee

At the beginning of this section we have proposed to take into account the space-time curvature in an action with a deformed  Casimir considering the substitution $k\rightarrow \bar{k}=\bar{e}k$ in the DDR, as this works for the transition from SR to GR. We have just seen that if $ \mathcal{C}(k)$ can be viewed as a function of the distance from the origin to a point $k$ of the momentum metric $g^k_{\mu\nu} (k)$,  $ \mathcal{C}(\bar{k})$ is the same function of the distance from $(x,0)$ to $(x,k)$ of the momentum metric $g_{\mu\nu} (x,k)$, since the new DLT leave the new DDR invariant, so it can be considered as (a function of) the squared distance calculated with the new metric. This is in accordance with our initial assumption of taking $ \mathcal{C}(\bar{k})$  as the DDR in presence of a curved spacetime.

\section{Cotangent bundle geometry}
\label{sec:cbg}
When we have considered the RDK as a way to go beyond SR, we have not taken into account its possible effects on the space-time metric. Since we were able to describe the main kinematical ingredients from a geometrical point of view, it is natural to seek for a geometrical description of spacetime which also allows us to deal with curved spacetimes. This leads to a geometrical construction in the whole phase space, leading to a metric depending on both space-time and momentum coordinates.

There are a lot of works in the literature studying the space-time consequences  of a velocity dependent metric~\cite{Papagiannopoulos:2017whb,Hohmann:2018rpp,Pfeifer:2019tyy,Hohmann:2019sni,Hohmann:2020yia,Heefer:2020hra,Triantafyllopoulos:2020vkx}, and in particular,  in the setting of LIV scenarios~\cite{Kostelecky:2011qz,Barcelo:2001cp,Weinfurtner:2006wt,Hasse:2019zqi,Stavrinos:2016xyg}. Most of them have been developed by considering Finsler geometries, formulated by Finsler in 1918~\cite{zbMATH02613491} (these geometries are a generalization of Riemannian spaces where the space-time metric can depend also on vectors of the tangent space). However, in those works the introduction of a velocity dependent metric is considered out of the DSR context since there is no mention to a DLT and DCL, hence precluding the possibility to have a relativity principle.

In the DSR framework, the starting point of most of the papers in the literature is also a DDR. However, there is a crucial difference between the LIV and DSR scenarios, since the latter implements a deformed Lorentz transformations (in the one-particle system) which makes the DDR invariant for different observers related by such transformation. The case of Finsler geometries in this context was considered for the cases of flat~\cite{Girelli:2006fw,Amelino-Camelia:2014rga} and curved spacetimes~\cite{Letizia:2016lew}, provoking a velocity dependence on the space-time metric. Besides Finsler geometries, which starts from a Lagrangian (in fact, Finsler geometries are particular realizations of Lagrange spaces~\cite{miron2001geometry}), there is another possible approach to define a deformed metric from the Hamiltonian. This leads to Hamilton geometry~\cite{miron2001geometry}, considered in~\cite{Barcaroli:2016yrl,Barcaroli:2015xda,Barcaroli:2017gvg}. In this kind of approach, the space-time metric depends on the phase-space coordinates (momentum and positions), instead of the tangent bundle coordinates (velocities and positions).  Both geometries are particular cases of geometries in the tangent and cotangent bundle respectively. 

Moreover, in~\cite{Rosati:2015pga} another way to consider possible phenomenology based on time delays in an expanding universe due to deformations of SR is explored. In that paper, it is studied both LIV and DSR scenarios, starting with a DDR and considering nontrivial translations. In order to do so, they considered the expansion of the universe by gluing slices of de Sitter spacetimes, finding difficult to formulate such study in a direct way. 

As we have mentioned, the DDR and the one-particle DLT are the only ingredients in all previous works. But as we discussed previously, there is a particular basis in  $\kappa$-Poincaré, the classical basis, in which the DDR and DLT are just the ones of SR; so, following the prescription used in these works, there would be no effect on the space-time metric.

Here we will study the case of  curved spacetime and momentum spaces by considering a geometry in the cotangent bundle known as generalized Hamilton geometries~\cite{miron2001geometry}, i.e., a geometrical structure for all the phase-space coordinates~\cite{Relancio:2020mpa,Relancio:2020zok,Pfeifer:2021tas,Relancio:2020rys}. As we will see, this is mandatory in order to make compatible a de Sitter momentum space and a generic space-time geometry. With our prescription, we find a nontrivial (momentum dependent) metric for whatever form of the DDR, being able to describe the  propagation of a free particle in the canonical variables. This differs from the perspective of~\cite{Rosati:2015pga}, where the considered metric for the spacetime is the one given by GR.

\subsection{Main properties of the geometry in the cotangent bundle}
\label{sec:hamilton-metric}

The aim of this subsection is to introduce the main ingredients and characteristics of a cotangent bundle geometry we will use in the following, and in particular, of the so-called generalized Hamilton spaces~\cite{miron2001geometry}. We start by making a short introduction to the cotangent bundle structure for the GR case in order to make the reader familiar with it.

We start by considering a base manifold $M$ with $n$  dimensions, which in GR plays the role of the curved spacetime in which particles move when $n=4$. One can construct from it the cotangent bundle manifold $T^*M$ with dimension $2n$, which is formed by  the base manifold (the spacetime) and the fibers (the momentum space). This allows us to describe geometrically all the phase space, which is very necessary for taking into account the momentum dependency of the space-time metric.

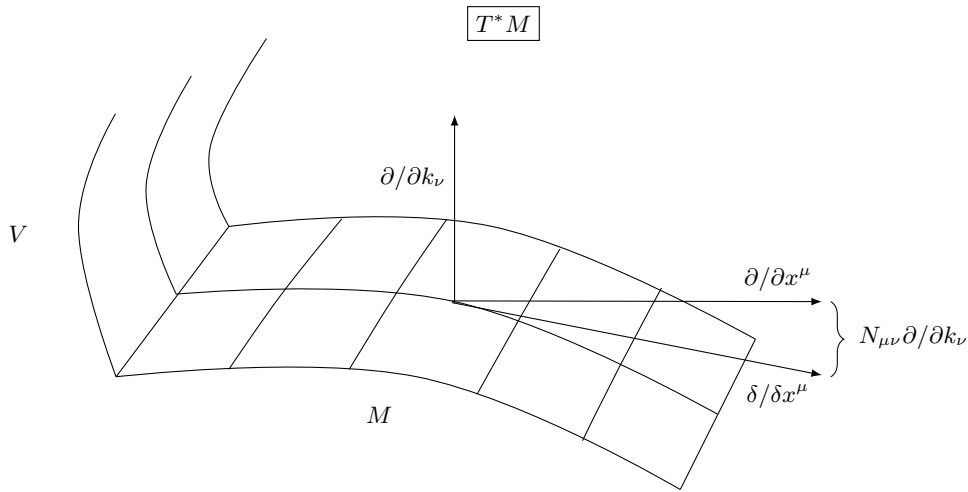
\begin{figure}
\centering
\begin{tikzpicture}
\draw (-4,1.5) node (v2) {} -- (-5.5,-0.5) node (v1) {};
\draw  plot[smooth, tension=.7] coordinates {(v1) (-1.5,-0.5) (2,-2)};
\draw  plot[smooth, tension=.7] coordinates {(v2)};
\draw  plot[smooth, tension=.7] coordinates {(v2) (-0.5,1.5) (3,0)};
\draw (2,-2) -- (3,0);
\draw  plot[smooth, tension=.7] coordinates {(-2.5,1.6) (-3.3,0.6) (-4,-0.4)};
\draw  plot[smooth, tension=.7] coordinates {(-1.1,1.6) (-1.75,0.65) (-2.4,-0.4)};
\draw  plot[smooth, tension=.7] coordinates {(0.4,1.2) (-0.15,0.25) (-0.7,-.73)};
\draw  plot[smooth, tension=.7] coordinates {(1.75,0.68) (1.2,-0.35) (0.7,-1.35)};
\draw  plot[smooth, tension=.7] coordinates {(-4.7,0.6) (-1,0.5) (2.5,-1)};
\node (v4) at (-1,0.36) {};
\node (v3) at (-1,3.1) {};
\node (v5) at (-1.55,2.15) {$\partial/\partial k_\nu$};
\draw [-latex](v4) -- (v3);
\node (v6) at (4,-0.5) {};
\node (v8) at (3.3,-0.75) {$\delta/\delta x^\mu$};
\node (v9) at (-1.15,0.51) {};
\draw [-latex](v9) -- (v6);
\node (v7) at (4,0.5) {};
\draw [-latex](v9) -- (v7);
\node at (3.3,0.8) {$\partial/\partial x^\mu$};
\draw [decorate,decoration={brace,amplitude=5pt,mirror,raise=0pt},yshift=0pt]
(4,-0.5) -- (4,0.5) node [font=\bfseries,midway,xshift=1.1cm,yshift=1pt]
{$N_{\mu \nu}\partial/\partial k_\nu$};
\node at (-2,-1) {$M$};
\node at (-6.8,1.4) {$V$};
\draw  plot[smooth, tension=.7] coordinates {(v1) (-6,1.5) (-5.5076,3)};
\draw  plot[smooth, tension=.7] coordinates {(-4.7,0.6) (-5.1,2.05) (-4.5,3.5)};
\draw  plot[smooth, tension=.7] coordinates {(v2) (-4.25,2.55) (-3.5,4)};
\node[draw,align=left] at (-0.35,4.2) {$T^*M$};
\end{tikzpicture}
\caption{Visualization of the cotangent structure.}
\label{figure_cotangent}
\end{figure}

We can associate  to each point $u \in T^*M$ on the cotangent bundle manifold, described by a point in phase space $(x,k)$, the fiber $V_u$, which is formed by  all the points with the same $x$ but with different $k$. We can then obtain the so-called vertical distribution, $V: u \in T^*M\rightarrow V_u \subset T_u T^*M$, with dimension $n$, which is generated by $\partial/\partial k$, being $T_u T^*M$ the tangent space of the manifold $T^*M$. Following the illustration of the cotangent bundle in Fig~\ref{figure_cotangent}, given a point on the base manifold one can construct the vertical distribution, or in other words, the fiber. Note that in the figure the fiber is unidimensional for the sake of simplicity, but in fact it has the same dimensions as the base manifold.

As it is shown in~\cite{miron2001geometry}, it is possible to define a  nonlinear connection $N$ (also called horizontal distribution), which is supplementary to the vertical distribution $V$, i.e., $ T_u T^*M=N_u\oplus V_u$, having also dimension $n$. This allows us to construct an adapted basis for the horizontal distribution
\begin{equation}
\frac{\delta}{\delta x^\mu}\, \doteq \,\frac{\partial}{\partial x^\mu}+N_{\nu\mu}(x,k)\frac{ \partial}{\partial k_\nu}\,,
\label{eq:delta_derivative}
\end{equation}  
where $N_{ \nu \mu}$ are the coefficients of the nonlinear connection. There are several possible choices of these coefficients but, as it is discussed in~\cite{miron2001geometry}, one and only one choice of nonlinear connection coefficients leads to a  metric compatible space-time affine connection  which is torsion free.
In GR, these coefficients can be written as 
\begin{equation}
N_{\mu\nu}(x,k)\, = \, k_\rho {\Gamma^\rho}_{\mu\nu}(x)\,,
\label{eq:nonlinear_connection}
\end{equation} 
where $ {\Gamma^\rho}_{\mu\nu}(x)$ is the affine connection. Therefore, these coefficients vanish when the metric is independent on the space-time coordinates. 

The difference between the horizontal distribution generated by $\delta/\delta x$ and the curves generated by $\partial/\partial x$ is represented in Fig~\ref{figure_cotangent}. The vector $\delta/\delta x$ provokes a motion not only on the base manifold (spacetime) but also along the fiber (momentum space). We can go back again to the GR example as an illustration: when following a geodesic, a particle changes not only its space-time coordinates but also its momentum, which is depicted by a movement along the fiber. This movement becomes nontrivial when the metric depends on all  phase-space coordinates, as we will see in the following. 

The line element in the cotangent bundle is~\cite{miron2001geometry}  
\begin{equation}
\mathcal{G}\,=\, g_{\mu\nu}(x,k) dx^\mu dx^\nu+g^{\mu\nu}(x,k) \delta k_\mu \delta k_\nu\,,
\label{eq:line_element_ps} 
\end{equation}
where 
\begin{equation}
\delta k_\mu \,=\, d k_\mu - N_{\nu\mu}(x,k)\,dx^\nu\,. 
\end{equation}

The \textit{d-curvature tensor} is defined as~\cite{miron2001geometry}
\begin{equation}
R_{\mu\nu\rho}(x,k)\,=\,\frac{\delta N_{\nu\mu}(x,k)}{\delta x^\rho}-\frac{\delta N_{\rho\mu}(x,k)}{\delta x^\nu}\,,
\label{eq:dtensor}
\end{equation} 
which represents the curvature of the phase space. In GR, it is proportional to the Riemann tensor 
\begin{equation}
R_{\mu\nu\rho}(x,k)\,=\,k_\sigma {R^{\sigma}}_{\mu\nu\rho}(x)\,.
\end{equation} 
It measures the integrability of spacetime, i.e., position space, as a subspace of the cotangent bundle, and is defined as the commutator between the horizontal vector fields
\begin{equation}
\left[ \frac{\delta}{\delta x^\mu}\,,\frac{\delta}{\delta x^\nu}\right]\,=\,\frac{\delta}{\delta x^\mu}\frac{\delta}{\delta x^\nu}-\frac{\delta}{\delta x^\nu}\frac{\delta}{\delta x^\mu}\,=\, R_{\rho\mu\nu}(x,k)\frac{\partial}{\partial k_\rho}\,.
\label{eq:commutator_deltas}
\end{equation}

Looking at the line element of Eq.~\eqref{eq:line_element_ps},  we can see that there are two preferred types of curves~\cite{miron2001geometry}: ones for which $dx^\mu$ is zero, giving the movement along a fiber, and another ones for which $\delta k_\mu$ is zero, leading to the space-time geodesics. A vertical path is characterized as a curve in the cotangent bundle with constant space-time coordinates and with the momentum satisfying the geodesic equation with the connection of the momentum space, i.e.
\begin{equation}
x^\mu\left(\tau\right)\,=\,x^\mu_0\,,\qquad \frac{d^2k_\mu}{d\tau^2}-{C^{\nu\sigma}}_\mu(x_0,k)\frac{dk_\nu}{d\tau}\frac{dk_\sigma}{d\tau}\,=\,0\,,
\end{equation} 
where 
\begin{equation}
{C^{\mu\nu}}_\rho(x,k)\,=\,-\frac{1}{2}g_{\rho\sigma}\left(\frac{\partial g^{\sigma\nu}(x,k)}{\partial k_ \mu}+\frac{\partial g^{\sigma\mu}(x,k)}{\partial k_ \nu}-\frac{\partial g^{\mu \nu}(x,k)}{\partial k_ \sigma}\right)\,,
\label{eq:affine_connection_p}
\end{equation}
is the affine connection in momentum space, while a horizontal curve will be determined by the geodesic motion in spacetime given by
\begin{equation}
\frac{d^2x^\mu}{d\tau^2}+{H^\mu}_{\nu\sigma}(x,k)\frac{dx^\nu}{d\tau}\frac{dx^\sigma}{d\tau}\,=\,0\,,
\label{eq:horizontal_geodesics_curve_definition}
\end{equation} 
and the change of momentum obtained from
\begin{equation}
\frac{\delta k_\lambda}{d \tau}\,=\,\frac{dk_\lambda}{d\tau}-N_{\sigma\lambda} (x,k)\frac{dx^\sigma}{d\tau}\,=\,0\,,
\label{eq:horizontal_momenta}
\end{equation} 
where 
\begin{equation}
{H^\rho}_{\mu\nu}(x,k)\,=\,\frac{1}{2}g^{\rho\sigma}(x,k)\left(\frac{\delta g_{\sigma\nu}(x,k)}{\delta x^\mu} +\frac{\delta g_{\sigma\mu}(x,k)}{\delta  x^\nu} -\frac{\delta g_{\mu\nu}(x,k)}{\delta x^\sigma} \right)\,,
\label{eq:affine_connection_st}
\end{equation} 
is the affine connection of spacetime. Again, $\tau$ plays the role of the proper time or the affine parametrization depending if one is considering a massive or a massless particle, respectively. 

It is possible to define in this scheme a covariant derivatives in space-time~\cite{miron2001geometry} 
\begin{equation}
\begin{split}
&T^{\alpha_1 \ldots\alpha_r}_{\beta_1\ldots\beta_s;\mu}(x,k)\,=\,\frac{\delta T^{\alpha_1 \ldots\alpha_r}_{\beta_1\ldots\beta_s}(x,k)}{\delta x^\mu}+T^{\lambda \alpha_2 \ldots\alpha_r}_{\beta_1\ldots\beta_s}(x,k){H^{\alpha_1}}_{\lambda \mu}(x,k)+\cdots+\\
&T^{\alpha_1 \ldots \lambda}_{\beta_1\ldots\beta_s}(x,k){H^{\alpha_r}}_{\lambda \mu}(x,k)-T^{\alpha_1 \ldots \alpha_r}_{\lambda \beta_2\ldots\beta_s}(x,k){H^{\lambda}}_{\beta_1 \mu}(x,k)-\cdots-T^{\alpha_1 \ldots \alpha_r}_{\beta_1\ldots \lambda}(x,k){H^{\lambda}}_{\beta_s \mu}(x,k)\,,
\label{eq:cov_dev_st}
\end{split}
\end{equation} 
and in momentum space
\begin{equation}
\begin{split}
&T^{\alpha_1 \ldots\alpha_r;\mu}_{\beta_1\ldots\beta_s}(x,k)\,=\,\frac{\partial T^{\alpha_1 \ldots\alpha_r}_{\beta_1\ldots\beta_s}(x,k)}{\partial k_\mu}+T^{\lambda \alpha_2 \ldots\alpha_r}_{\beta_1\ldots\beta_s}(x,k){C^{\alpha_1\mu}}_{\lambda}(x,k)+\cdots+\\
&T^{\alpha_1 \ldots \lambda}_{\beta_1\ldots\beta_s}(x,k){C^{\alpha_r \mu}}_{\lambda }(x,k)-T^{\alpha_1 \ldots \alpha_r}_{\lambda \beta_2\ldots\beta_s}(x,k){C^{\lambda \mu}}_{\beta_1 }(x,k)-\cdots-T^{\alpha_1 \ldots \alpha_r}_{\beta_1\ldots \lambda}(x,k){C^{\lambda \mu}}_{\beta_s}(x,k)\,,
\label{eq:cov_dev_ms}
\end{split}
\end{equation} 
Also, in~\cite{miron2001geometry} it is discussed the fact that, for a given metric, it exists always a symmetric non-linear connection leading to the affine connections in space-time and in momentum spaces making that both covariant derivatives of the metric vanishes:
\begin{equation}
g_{\mu\nu;\rho}\,=\,g_{\mu\nu}^{\,\,\,\,\,\,;\rho}\,=\,0\,.
\label{eq:covariant_derivative_2}
\end{equation}

\subsection{Relationship between metric and the Casimir}
\label{subsec_action_metric}
In this subsection we will see a simple way to obtain the Casimir, defined as a function of the square of the distance in momentum space from the origin to a point $k$, from a metric.  

In~\cite{Bhattacharya2012RelationshipBG} it is showed that there is a simple equation relating   the distance $D(0,k)$  from a fixed point $0$ to $k$ of a Riemannian manifold  and its metric. In~\cite{Relancio:2020zok} we obtain, starting from that relationship,
 \begin{equation}
\frac{\partial D(0,k)}{\partial k_ \mu}g^k_{\mu\nu}(k) \frac{\partial D(0,k)}{\partial k_ \nu}\,=\,1\,.
\end{equation}
Moreover, this property is also checked in Ch.~3 of~\cite{Petersen2006} for the Minkowski space (inside the light cone and extended on the light cone by continuity) and so, by the Whitney embedding theorem~\cite{Burns1985}, valid for any Riemannian manifold of dimension $n$, since they can be embedded in a Minkowski space of at most dimension $2n+1$. Through this property, we can establish a direct relationship between the metric and the Casimir defined as the square of the distance
 \begin{equation}
\frac{\partial  \mathcal{C}(k)}{\partial k_ \mu}g^k_{\mu\nu}(k) \frac{\partial  \mathcal{C}(k)}{\partial k_ \nu}\,=\,4  \mathcal{C}(k)\,.
\label{eq:casimir_definition}
\end{equation}
It is then trivial to obtain from this expression any function of the squared distance. 

One can also arrive to the same relation  for the generalization proposed of the cotangent bundle metric.   Therefore, Eq.~\eqref{eq:casimir_definition} is generalized to
 \begin{equation}
\frac{\partial  \mathcal{C}(\bar{k})}{\partial \bar{k}_ \mu}g^{\bar{k}}_{\mu\nu}(\bar{k}) \frac{\partial  \mathcal{C}(\bar{k})}{\partial \bar{k}_ \nu}\,=\,4  \mathcal{C}(\bar{k})\,=\,\frac{\partial  \mathcal{C}(\bar{k})}{\partial k_ \mu}g_{\mu\nu}(x,k) \frac{\partial  \mathcal{C}(\bar{k})}{\partial k_\nu}\,.
\label{eq:casimir_definition_cst}
\end{equation}
We see that the same relation found for the flat space-time case holds also for curved spacetime. 

\subsection{Geodesics from Hamilton equations}
\label{sub:geo}
In Hamilton spaces,  the nonlinear connection can be obtained from the metric (and then from the Hamiltonian, since they are also related in a simple way) in a straightforward fashion~\cite{miron2001geometry}. However, this is not the case for a generalized Hamilton space $GH^n$, as it is discussed in~\cite{miron2001geometry}: ``In general, we cannot determine a nonlinear connection from the fundamental tensor $g^{ij}$ of the space $GH^n$. Therefore, we study the N-linear connections compatible with $g^{ij}$, $N$ being a priori given'' (see the beginning of page 121).  Moreover, there is neither a connection between a Hamiltonian and the cotangent bundle metric. In~\cite{Relancio:2020rys}, we proposed a way to relate these three ingredients for a generalized Hamilton space.  

We start by discussing the geodesic equation via Hamilton equations. We want to recover~\eqref{eq:horizontal_geodesics_curve_definition}  starting from a deformed Casimir. This imposes a  relationship between the nonlinear coefficients and the space-time affine connection.  For that aim, we compute the equations of motion from a Hamiltonian $\mathcal{C}$:
\begin{equation}
\frac{dx^\mu}{d\tau}\,=\,\mathcal{N}\lbrace{\mathcal{C} (x,k),x^\mu\rbrace}\,=\,\mathcal{N}\frac{\partial \mathcal{C} (x,k)}{\partial k_\mu}\,,\quad\frac{dk_\mu}{d\tau}\,=\,\mathcal{N}\lbrace{\mathcal{C} (x,k),k_\mu\rbrace}\,=\,-\mathcal{N}\frac{\partial \mathcal{C} (x,k)}{\partial x^\mu}\,,
\label{eq:H-eqs}
\end{equation}
where again $\mathcal{N}$ is a Lagrange multiplier, being $1/2m$ or $1$ for massive and massless particles respectively, and we used the Poisson bracket 
\begin{equation}
\lbrace{f,g\rbrace}\,=\,\frac{\partial f}{\partial k_\rho}\frac{\partial g}{\partial x^\rho}-\frac{\partial g}{\partial k_\rho}\frac{\partial f}{\partial x^\rho}\,.
\label{eq:pb}
\end{equation}

For a horizontal curve Eq.~\eqref{eq:horizontal_momenta} holds, so substituting Eq.~\eqref{eq:H-eqs} in Eq.~\eqref{eq:horizontal_momenta} we obtain 
\begin{equation}
 \frac{\delta k_\lambda}{d \tau}\,=\,\frac{dk_\lambda}{d\tau}-N_{\sigma\lambda} (x,k)\frac{dx^\sigma}{d\tau}\,=\,-\mathcal{N}\left(\frac{\partial \mathcal{C} (x,k)}{\partial x^\lambda}+N_{\sigma\lambda}\frac{\partial \mathcal{C} (x,k)}{\partial k_\sigma} \right)\,=\,0\,,
\label{eq:derivative_casimir}
 \end{equation}
which implies
\begin{equation}
 \frac{\delta \mathcal{C} (x,k)}{\delta x^\mu}\,=\,\mathcal{C} _{;\mu}(x,k)\,=\,0\,.
\label{eq:delta_casimir}
\end{equation}
This was obtained in~\cite{Barcaroli:2015xda} in the Hamilton geometry context. This can be easily understood from the fact that since the Casimir is a constant along geodesics, its covariant derivative should vanish. It is important to note that the usual covariant derivative of GR (which for a function reduces to the partial with respect to the space-time coordinates) applied to the Casimir is not zero. However, as showed in~\cite{Relancio:2020rys}, Eq.~\eqref{eq:delta_casimir} is satisfied in GR for the Casimir $\mathcal{C} (x,k)=k_\mu g^{\mu\nu}(x)k_\nu$. 

Turning back to our discussion, we can compute the second derivative of the position finding
\begin{equation}
\frac{d^2x^\mu}{d\tau^2}\,=\, \mathcal{N}\frac{d}{d\tau}\frac{\partial \mathcal{C} (x,k)}{\partial k_\mu}\,=\, \mathcal{N}\left(\frac{\partial^2 \mathcal{C} (x,k)}{\partial k_\mu \partial x^\rho} \frac{dx^\rho}{d\tau}+\frac{\partial^2 \mathcal{C} (x,k)}{\partial k_\mu \partial k_\sigma} \frac{dk_\sigma}{d\tau}\right)\,.
\label{eq:geo_1}
\end{equation}
We can rewrite the first term in the right-hand side of the previous equation using Eq.~\eqref{eq:derivative_casimir}, obtaining
\begin{equation}
\begin{split}
\frac{\partial^2 \mathcal{C} (x,k)}{\partial k_\mu \partial x^\rho}\,=&\, \frac{\partial }{\partial k_\mu} \frac{\partial \mathcal{C} (x,k)}{\partial x^\rho}\,=\, - \frac{\partial }{\partial k_\mu} N_{\rho \sigma}(x,k) \frac{\partial \mathcal{C} (x,k)}{\partial k_\sigma}\,=\,\\
&- \frac{\partial N_{\rho \sigma}(x,k)}{\partial k_\mu}  \frac{\partial \mathcal{C} (x,k)}{\partial k_\sigma}- N_{\rho \sigma}(x,k) \frac{\partial^2 \mathcal{C} (x,k)}{ \partial k_\mu \partial k_\sigma}\,.
\end{split}
\end{equation}

Therefore, using ~\eqref{eq:horizontal_momenta}, Eq.~\eqref{eq:geo_1} becomes
\begin{equation}
\begin{split}
\frac{d^2x^\mu}{d\tau^2}\,=&\,  \mathcal{N}\left(- \frac{\partial N_{\rho \sigma}(x,k)}{\partial k_\mu}  \frac{\partial \mathcal{C} (x,k)}{\partial k_\sigma}- N_{\rho \sigma}(x,k) \frac{\partial^2 \mathcal{C} (x,k)}{\partial k_\mu\partial k_\sigma}+\frac{\partial^2 \mathcal{C} (x,k)}{\partial k_\mu \partial k_\sigma} N_{\rho \sigma} (x,k)\right) \frac{dx^\rho}{d\tau}\\
\,=&\,-  \mathcal{N} \frac{\partial N_{\rho \sigma}(x,k)}{\partial k_\mu}  \frac{\partial \mathcal{C} (x,k)}{\partial k_\sigma} \frac{dx^\rho}{d\tau}\,.
\end{split}
\end{equation}

Finally, using the first expression of Eq.\eqref{eq:H-eqs} one finds 
\begin{equation}
\frac{d^2x^\mu}{d\tau^2}+{H^\mu}_{\nu\sigma}(x,k)\frac{dx^\nu}{d\tau}\frac{dx^\sigma}{d\tau}\,=\,0\,,
\label{eq:horizontal_geodesics}
\end{equation}
where 
\begin{equation}
{H^\mu}_{\nu\sigma}(x,k)\,=\,\frac{\partial N_{\nu\sigma}(x,k)}{\partial k_\mu}\,.
\label{eq:affine_connection_n}
\end{equation}
For the Casimir of GR, which is a quadratic expression of the momentum, one gets ${H^\mu}_{\nu\sigma}(x,k)={\Gamma^\rho}_{\mu\nu}(x)$, so the nonlinear connection is given by Eq.~\eqref{eq:nonlinear_connection}. By imposing that the same equations of motions are obtained from the Hamilton equations and from the geodesic equation of the metric, one obtains the relationship~\eqref{eq:affine_connection_n} between the nonlinear coefficients and the space-time affine connection.

\subsection{Conservation of the Casimir along an horizontal path}
\label{sec:casimir_horizontal}

We have assumed along this review that a function of the squared of the distance in momentum space is the Casimir. Here indeed we show that this distance is conserved along a horizontal path, and then, one can consider a function of such distance as the Casimir~\cite{Relancio:2020rys}.

An infinitesimal displacement $\Delta x^\mu$ along a horizontal curve changes the phase-space coordinates as 
\begin{equation}
x^{\prime \mu}\,=\,x^\mu + \Delta x^\mu\,,\qquad k^{\prime }_\nu\,=\,k_\nu + N_{\mu \nu} (x,k)\,\Delta x^\mu\,.
\label{eq:variation_horizontal}
\end{equation}
Let us assume that the momentum line element is conserved along a horizontal curve. Then, the distances between $(x,0)$ to $(x,k)$ and  $(x^\prime,0)$ to  $(x^\prime,k^\prime)$ are the same, which implies that the following equation must hold
\begin{equation}
dk^\prime_\mu g^{\mu \nu} (x^\prime,k^\prime)dk^\prime_\nu\,=\,dk_\rho g^{\rho \sigma} (x,k) dk_\sigma\,.
\label{eq:momentum_isometry}
\end{equation}
This  expression can be rewritten as 
\begin{equation}
g_{\mu \nu} (x^\prime,k^\prime)\,=\, \frac{\partial k^\prime_\mu }{\partial k_\rho} g_{\rho \sigma} (x,k)\frac{\partial k^\prime_\nu }{\partial k_\sigma} \,,
\label{eq:momentum_isometry2}
\end{equation}
and using Eq.~\eqref{eq:variation_horizontal} we finally obtain
\begin{equation}
\begin{split}
\frac{\delta g_{\mu \nu}(x,k)}{\delta x^\rho}-\frac{\partial N_{\mu \rho}(x,k)}{\partial k_\lambda} g_{\lambda \nu}(x,k)-\frac{\partial N_{\nu \rho}(x,k)}{\partial k_\lambda} g_{\lambda \mu}(x,k)\,=\,&\\
\frac{\delta g_{\mu \nu}(x,k)}{\delta x^\rho}-{H^\lambda}_{\mu \rho}(x,k) g_{\lambda \nu}(x,k)-{H^\lambda}_{\nu \rho}(x,k) g_{\lambda \mu}(x,k)\,=\,&0\,,
\end{split}
\label{eq:covariant_derivative}
\end{equation}
where in the second equation we have used the relation~\eqref{eq:affine_connection_n} between the space-time affine connection and the nonlinear connection, and in the last step that the space-time covariant derivative of the metric vanishes~\eqref{eq:covariant_derivative_2}.

We can summarize here the results of these two last subsections. We have gone  beyond the understanding of~\cite{miron2001geometry} for generalized Hamilton spaces by 
defining, in an unequivocal way, all the cotangent bundle ingredients describing particle trajectories starting from a metric: the space-time affine connection, the nonlinear connection, and the Casimir, can be determined using  Eqs.~\eqref{eq:affine_connection_st},\eqref{eq:delta_casimir},\eqref{eq:affine_connection_n} and the fact that the Casimir is a function of the squared of the distance in momentum space.

\subsection{Covariant derivative along a curve}
In this subsection we revise the concept of covariant derivative along a curve in the cotangent bundle geometry context~\cite{Relancio:2020rys}. Let us start by considering that, as in GR~\cite{Weinberg:1972kfs}, a tensor $A^\mu(\tau)$ transforms under a diffeomorphisms as 
\begin{equation}
A^{\prime \mu}(\tau)\,=\, \frac{\partial x^{\prime \mu}}{\partial x^\nu}A^\nu(\tau)\,,
\end{equation}
so
\begin{equation}
\frac{d A^{\prime \mu}(\tau)}{d\tau}\,=\, \frac{\partial x^{\prime \mu}}{\partial x^\nu}\frac{d A^\nu(\tau)}{d\tau}+\frac{\partial^2 x^{\prime \mu}}{\partial x^\nu\partial x^\rho}\frac{dx^\rho}{d\tau}A^\nu(\tau)\,.
\end{equation}

Taking into  account that the modified affine connection~\eqref{eq:affine_connection_st} transforms exactly as the usual connection of GR~\cite{miron2001geometry}, we can then define a covariant derivative along a curve $x^\mu(\tau)$ as 
\begin{equation}
\frac{D A^{\mu}(\tau)}{D\tau}\,\coloneqq\,\frac{d A^{\mu}(\tau)}{d\tau}+{H^\mu}_{\nu \rho}(x,k)\frac{dx^\rho}{d\tau}A^\nu(\tau) \,.
\label{cov_der_curve}
\end{equation}
An analogous expression for a Finsler space was used in~\cite{Rund2012}. Therefore we can write the geodesic equation as in GR
 \begin{equation}
 \begin{split}
\frac{D u^{\mu}}{D\tau} &\,=\,\frac{\partial u^\mu }{\partial x^\nu}\frac{d x^\nu}{d \tau}+\frac{\partial u^\mu }{\partial k_\rho }\frac{d k_\rho}{d \tau}+{H^{\mu}}_{\nu \sigma}(x,k) u^\nu u^\sigma\\
&\,=\,\frac{\delta u^\mu }{\delta x^\nu} u^\nu+{H^{\mu}}_{\nu \sigma}(x,k) u^\nu u^\sigma \,=\,u^\mu_{;\nu}u^\nu\,=\,0\,, 
 \end{split}
\label{eq:geodesic_cov_curve}
\end{equation}
being $u^\mu=dx^\mu/d\tau$, and where we have used that for a horizontal curve~\eqref{eq:horizontal_momenta} holds.

\subsection{Lie derivative}
In this subsection we derive the modified Lie derivative for a contravariant vector since we will use this result in the following~\cite{Relancio:2020zok,Relancio:2020rys}. We can express the variation of the coordinates $x^\alpha$ along a vector field $\chi^\alpha(x)$ as 
\begin{equation}
x^{\prime \alpha}(x)\,=\,x^\alpha+\chi^\alpha(x) \Delta\lambda\,,
\end{equation}
where $\lambda$ is he infinitesimal variation parameter. This variation of $x^\alpha$ reflects on $k_\alpha$ in the following way
\begin{equation}
k^\prime_{\alpha}\,=\,k_\beta \frac{\partial x^\beta}{\partial x^{\prime \alpha}}\,=\,k_\alpha-\frac{\partial\chi^\beta(x)}{\partial x^\alpha}k_\beta \Delta\lambda\,,
\label{eq:x_variation}
\end{equation}
since $k$ transforms as a covector. The general variation of a vector field $u^\alpha\left(x,k\right)$ will then be
\begin{equation}
\begin{split}
\Delta u^\alpha(x,k)\,=\,&\frac{\partial u^\alpha(x,k)}{\partial x^\beta} \Delta x^\beta+\frac{\partial u^\alpha(x,k)}{\partial k_\beta} \Delta k_\beta\\
\,=\,&\frac{\partial u^\alpha(x,k)}{\partial x^\beta}\chi^\beta (x) \Delta\lambda-\frac{\partial u^\alpha(x,k)}{\partial k_\beta}\frac{\partial\chi^\gamma (x) }{\partial x^\beta}k_\gamma \Delta\lambda \,.
\end{split}
\label{eq:vector_variation}
\end{equation}

Therefore, the Lie derivative is defined as 
\begin{equation}
\begin{split}
\mathcal{L}_\chi u^\mu(x,k)\,=\, &\frac{u^{\prime \mu}(x^\prime,k^\prime)-u^\mu(x^\prime,k^\prime)}{\Delta \lambda}\\
\,=\,&\chi^\nu(x) \frac{\partial u^\mu (x,k)}{\partial x^\nu}-u^\nu (x,k)\frac{\partial \chi^\mu (x)}{\partial x^\nu}-\frac{\partial u^\mu (x,k)}{\partial k_\alpha}\frac{\partial \chi^\gamma(x)}{\partial x^\alpha} k_\gamma\,.
\end{split}
\label{eq:lie_def}
\end{equation}
Using the definition~\eqref{eq:cov_dev_st} for the covariant derivative and taking into account that the vector field $\chi(x)$ does not depend on the momentum, one can finally write 
\begin{equation}
\begin{split}
\mathcal{L}_\chi u^\mu(x,k)\,=\,&\chi^\nu(x) u^\mu_{\,;\nu}(x,k)-u^\nu(x,k) \chi^\mu_{\,;\nu}(x)-\\
&\frac{\partial u^\mu(x,k)}{\partial k_\alpha} \frac{\partial \chi^\gamma(x)}{\partial x^\alpha} k_\gamma -\frac{\partial u^\mu(x,k)}{\partial k_\alpha}N_{\alpha \gamma}(x,k)\chi^\gamma(x)\,.
\end{split}
\label{eq:lie_vec}
\end{equation}

It is easy to see that in the limit of a $\Lambda$ independent vector field $u^\mu(x)$ the usual expression for the Lie derivative is recovered. 

\section{About the Einstein's equations and the ``physical''  basis}
\label{sec:einstein}
We can now go back to the question discussed in the introduction about the possibility that different basis of the kinematics, or in geometrical language, different momentum coordinates describing a de Sitter space, could represent different physics. In Sec.~\ref{sec:construction_metric_ps} we have said that, in the particular way in which we construct the metric in the cotangent bundle which allows us to define a RDK, the metric is invariant under diffeomorphisms. As we will discuss in this section, indeed it is not invariant under momentum change of coordinates. This means that our scheme makes a distinction between different bases. In this section, we address this issue by considering the Einstein's equations in the cotangent bundle framework~\cite{miron2001geometry}.  

\subsection{Definition of curvature tensors}

From the commutator of  two space-time covariant derivatives of a vector field $X^\mu (x,k)$ (in absence of torsion) one can define the space-time curvature tensor~\cite{miron2001geometry}
 \begin{equation}
X^\mu (x,k)_{;\nu;\rho}-X^\mu (x,k)_{;\rho;\nu}\,=\,X^\lambda (x,k) {R^{*\mu}}_{\lambda \nu\rho}(x,k)-X^\mu (x,k)^{;\lambda} R_{\lambda \nu\rho}(x,k) \,, 
\label{commutator_cov_der2}
\end{equation}
where $R_{\lambda \nu\rho}(x,k)$ is the \textit{d-curvature tensor} tensor defined in Eq.~\eqref{eq:dtensor},
 \begin{equation}
{R^{*\mu}}_{\lambda \nu\rho}(x,k)\,=\,{R^{\mu}}_{\lambda \nu\rho}(x,k)+{C^{\mu \sigma}}_\lambda (x,k)R_{\sigma \nu\rho}(x,k)\,,
\label{eq:riemann_wrong}
\end{equation}
and
 \begin{equation}
 \begin{split}
     {R^\mu}_{\nu \rho\sigma}(x,k)\,=\,&\frac{\delta {H^\mu}_{\nu \rho}(x,k)}{\delta x^\sigma}-\frac{\delta {H^\mu}_{\nu \sigma}(x,k)}{\delta x^\rho}\\
     &+{H^\lambda}_{\nu \rho}(x,k){H^\mu}_{\lambda \sigma}(x,k)-{H^\lambda}_{\nu \sigma}(x,k){H^\mu}_{\lambda \rho}(x,k)\,.
 \end{split}
\label{riemann_st}
\end{equation}

In a similar way, from the commutator  of two momentum covariant derivatives one can define   the curvature tensor in momentum space~\cite{miron2001geometry}
 \begin{equation}
X^{\mu;\nu;\rho} (x,k)-X^{\mu;\rho;\nu}(x,k)\,=\,X^\lambda (x,k) {S^{\mu\nu\rho}}_{\lambda }(x,k) \,, 
\label{commutator_cov_der2_momentum}
\end{equation}
where
\begin{equation}
\begin{split}
{S^{\mu\nu\rho}}_{\sigma}(x,k)\,=\,& \frac{\partial {C^{\mu\nu}}_\sigma(x,k)}{\partial k_\rho}-\frac{\partial {C^{\mu\rho}}_\sigma(x,k)}{\partial k_\nu}\\
&+{C^{\lambda\nu}}_\sigma(x,k){C^{\mu\rho}}_\lambda(x,k)-{C^{\lambda\rho}}_\sigma(x,k){C^{\mu\nu}}_\lambda(x,k)\,.
\end{split}
\label{eq:Riemann_p}
\end{equation}

Moreover, there is another curvature tensor associated to the entanglement of momentum space and spacetime. This can be computed from the commutator between spacetime and momentum space~\cite{miron2001geometry}
 \begin{equation}
 \begin{split}
X^{\mu\,\,\,;\rho}_{\,\,;\nu} (x,k)-X^{\mu;\rho}_{\,\,\,\,\,\,\,;\nu}(x,k)\,=\,&X^\lambda (x,k) {P^{\mu\rho}}_{\lambda \nu}(x,k)\\
&- X^{\mu}_{\,\,;\lambda}(x,k){C^{\lambda\rho}}_\nu(x,k)-X^{\mu\,\,\,;\lambda}(x,k){P^\rho}_{\lambda\nu}(x,k)\,, 
 \end{split}
\label{commutator_cov_der2_momentum_spacetime}
\end{equation}
where
\begin{equation} 
{P^{\mu\rho}}_{\lambda \nu}(x,k)\,=\, \frac{\partial {H^\mu}_{\lambda\nu}(x,k)}{\partial k_\rho}-{C^{\mu\rho}}_{\lambda\,\,\,;\nu}(x,k)+{C^{\mu\lambda}}_{\nu}(x,k){P^\rho}_{\lambda\nu}(x,k)\,,
\label{eq:Riemann_sp}
\end{equation} 
being 
\begin{equation} 
{P^\rho}_{\lambda\nu}(x,k)\,=\,{H^\rho}_{\lambda\nu}(x,k)-\frac{\partial N_{\nu\lambda}(x,k)}{\partial k_\rho}\,.
\label{eq:P_tensor}
\end{equation}

\subsection{Privileged basis from Einstein's equations}
Following~\cite{miron2001geometry}, one can construct the Einstein's equations from the Riemann tensor of Eq.~\eqref{eq:riemann_wrong}
 \begin{equation}
R^*_{\mu \nu}(x,k)-\frac{1}{2}g_{\mu \nu}(x,k) R^*(x,k)\,=\,8\pi T_{\mu\nu}(x,k) \,, 
\label{eq:einstein_tensor_wrong}
\end{equation}
where   
 \begin{equation}
T_{\mu \nu}(x,k)\,=\, \frac{\delta S_m}{\delta g^{\mu\nu}(x,k)}\,,
\label{eq:energy_momentum_action}
\end{equation}
is the variation of the matter term in the Einstein-Hilbert action, and
 \begin{equation}
R^*_{\mu \nu}(x,k)\,=\, {R^{*\lambda}}_{\mu \nu\lambda }(x,k)\,,\qquad R^*(x,k)\,=\,R^*_{\mu \nu} (x,k)g^{\mu \nu}(x,k)\,.
\label{eq:ricci_scalar_wrong}
\end{equation}
In an analogous fashion,  the stress-energy tensor in momentum space can be defined as~\cite{miron2001geometry}
 \begin{equation}
S^{\mu \nu}(x,k)-\frac{1}{2}g^{\mu \nu}(x,k)S \,=\,8\pi T_{(k)}^{\mu\nu}(x,k)\,,
\label{eq:einstein_eqs_p}
\end{equation} 
and the stress-energy tensors that intertwines spacetime and momentum space
 \begin{equation}
 \begin{split}
g_{\rho\sigma}(x,k){P^{\rho\sigma}}_{\mu \nu}(x,k)\,=\,& P^{(1)}_{\mu \nu}(x,k)\,=\, \kappa T^{(1)}_{\mu \nu}(x,k)\,\\ g_{\nu\rho}(x,k){P^{\rho\sigma}}_{\mu \sigma}(x,k)\,=\,& P^{(2)}_{\mu \nu}(x,k)\,=\,- \kappa T^{(2)}_{\mu \nu}(x,k)\,.
 \end{split}
\label{eq:einstein_eqs_ps}
\end{equation}

By imposing the conservation of energy momentum tensor in spacetime one finds that ${P^{\mu\rho}}_{\lambda \nu}=0$~\cite{miron2001geometry}. This solves the problematic issue about what sources could generate such tensors and its physical interpretation. 

Since we want to describe a RDK from the geometrical ingredients of the geometry, we need the momentum space  to be   maximally symmetric. This implies  the scalar of curvature in momentum space to be constant $S=12/\Lambda^2$, and that indeed 
\begin{equation}
S_{\rho\sigma\mu\nu}\,\propto \, g_{\rho\mu}g_{\sigma\nu}-g_{\rho\nu}g_{\sigma\mu}\,.
\end{equation}
Then, the Einstein's equations for momentum space are given by~\cite{miron2001geometry} 
 \begin{equation}
S^{\mu \nu}(x,k)-\frac{1}{2}g^{\mu \nu}(x,k)S +\frac{6}{\Lambda^2}g^{\mu \nu}(x,k)\,=\, 0\,,
\label{eq:einstein_eqs_p2}
\end{equation}
where $S^{\mu \nu}=S^{\lambda \mu \rho \nu}g_{\lambda\rho}$ and $S=S^{\mu \nu}g_{\mu \nu}$. These Einstein's equations represents an empty universe with a cosmological constant (in momentum space).

By imposing  ${P^{\mu\rho}}_{\lambda \nu}=0$ and~\eqref{eq:einstein_eqs_p} with our particular prescription for constructing the cotangent bundle  metric~\eqref{eq:cotangent_metric_tetrads}, we find that one and only one momentum basis (coordinates of a de Sitter momentum space) satisfies this property, which is  conformally flat~\cite{Relancio:2020rys} 
\begin{equation}
g_{\mu \nu}(x,k)\,=\,\eta_{\mu\nu}\left(1-\frac{k_0^2-\vec{k}^2}{4\Lambda^2}\right)^2\,.
\label{eq:physical_metric}
\end{equation}
This metric was used for first time in~\cite{Bicak:2005yt} and in the context of DSR in~\cite{Mignemi:2008fj}\footnote{In~\cite{Relancio:2021asx} it was also pointed out the special role of this basis from completely different arguments.}. Since tensors are not invariant  under momentum diffeomorphisms, only for this particular momentum coordinates the condition ${P^{\mu\rho}}_{\lambda \nu}$ will be satisfied.

We can now see what are the DDR and DCL of the $\kappa$-Poincaré kinematics obtained from this conformally flat metric. We obtain that DDR computed from Eq.~\eqref{eq:casimir_definition} is
\begin{equation}
\mathcal{C}(k)\,=\,4 \Lambda^2 \arccoth^2\left(\frac{2 \Lambda}{\sqrt{k_0^2-\vec{k}^2}}\right)\,,
\label{eq:casimir_physical}
\end{equation}
and that the DCL obtained from Eq.~\eqref{T,J}, up to second order, is 
\begin{equation}
\begin{split}
\left(p\oplus q\right)_0\,=\,&p_0+q_0+\frac{\vec{p}\cdot \vec{q}}{\Lambda}-\frac{p_0 q_0 \left(p_0+q_0\right)}{4\Lambda^2}+\frac{\vec{p}\cdot \vec{q} \left(q_0-p_0\right)}{2\Lambda^2}+\frac{3\vec{p}^2 q_0}{4\Lambda^2}-\frac{\vec{q}^2 p_0}{4\Lambda^2}\,,\\
\left(p\oplus q\right)_i\,=\,&p_i+q_i+\frac{p_i q_0}{\Lambda}+p_i\left(\frac{q_0^2}{4\Lambda^2}-\frac{p_0 q_0}{2\Lambda^2}+\frac{\vec{p}\cdot \vec{q}}{2\Lambda^2}-\frac{\vec{q}^2}{4\Lambda^2}\right)\\
&+q_i\left(\frac{\vec{p}^2}{4\Lambda^2}-\frac{p_0^2}{4\Lambda^2}-\frac{p_0 q_0}{2\Lambda^2}+\frac{\vec{p}\cdot \vec{q}}{2\Lambda^2}\right)\,.
\end{split}
\end{equation}

For this metric it is easy to check that  the nonlinear connection is given by Eq.~\eqref{eq:nonlinear_connection}. We can see that Eq.~\eqref{eq:delta_casimir} is automatically satisfied for the nonlinear coefficients of~\eqref{eq:nonlinear_connection}. We start by writing~\eqref{eq:delta_casimir} 
\begin{equation}
\begin{split}
    \frac{\delta \mathcal{C} (x,k)}{\delta x^\mu}\,=\,&\mathcal{N}\left( \frac{\partial \mathcal{C} (x,k)}{\partial x^\mu}+N_{\mu\nu}(x,k)\frac{\partial \mathcal{C} (x,k)}{\partial k_\nu}\right)\\
    \,=\,&\mathcal{N}\frac{\partial \mathcal{C} (x,k)}{\partial \bar{k}_\rho}\left(\frac{\partial \bar{k}_\rho}{\partial x^\mu}+N_{\mu\nu}(x,k)\frac{\partial \bar{k}_\rho}{\partial k_\nu}\right) \,,
\end{split}
\label{eq:casimir_delta_2}
\end{equation}
where we have used the fact that the Casimir depends on the phase-space coordinates through $\bar{k}$. Now, we can realize that, for our case of study, the Casimir~\eqref{eq:casimir_physical} is a function of $k^2$. Therefore, the Casimir for a curved spacetime is the same function of $\bar{k}^2$ (as we have explained in Sec.~\ref{sec:rdkcst}). We can then rewrite Eq.~\eqref{eq:casimir_delta_2} as 
\begin{equation}
\begin{split}
\frac{\partial \mathcal{C} (x,k)}{\partial \bar{k}^2}\eta^{\rho \sigma}\bar{k}_\sigma\left(\frac{\partial \bar{k}_\rho}{\partial x^\mu}+N_{\mu\nu}(x,k)\frac{\partial \bar{k}_\rho}{\partial k_\nu}\right)\,=\,\\
\frac{\partial \mathcal{C} (x,k)}{\partial \bar{k}^2}\left(\frac{1}{2}k_\lambda \frac{\partial g^{\lambda\sigma}(x)}{\partial x^\mu}k_\sigma+ N_{\mu \rho}(x,k)g^{\rho \sigma}(x)k_\sigma\right)\,=\,&0 \,,
\end{split}
\end{equation}
since the term in parenthesis vanishes when the nonlinear connection is defined by Eq.~\eqref{eq:nonlinear_connection}~\cite{Relancio:2020rys}.  This means that the nonlinear connection~\eqref{eq:nonlinear_connection} is compatible with Eq.~\eqref{eq:delta_casimir}, and then with the metric in the cotangent space~\eqref{eq:metric_cotangent_definition}. Even more, using Eq.~\eqref{eq:affine_connection_n}, one can see that the space-time affine connection is also the same one of GR. 

We want the same energy-momentum tensor to be the same of the one of GR in this scheme, and then, to be momentum independent. However, we can see that  the left hand side of the  definition of the Einstein's equations~\eqref{eq:einstein_tensor_wrong} depends on the momentum due to the last term of Eq.~\eqref{eq:riemann_wrong}. This points out to consider as the Riemann tensor for constructing the Einstein's equations, instead of Eq.~\eqref{eq:riemann_wrong}, the one of  Eq.~\eqref{riemann_st}. In this case, we have 
 \begin{equation}
R_{\mu \nu}(x)-\frac{1}{2}g_{\mu \nu}(x,k)R(x,k) \,=\,8\pi T_{\mu \nu}(x)\,,
\label{eq:einstein_eqs}
\end{equation}
where
 \begin{equation}
R_{\mu \nu}(x)\,=\, R^{\lambda}_{\mu \nu\lambda }(x)\,,\qquad R(x,k)\,=\,R_{\mu \nu} (x)g^{\mu \nu}(x,k)\,.
\label{eq:ricci_scalar}
\end{equation}
We can note two important facts. The Riemann and Ricci tensors do not depend on $k$ due to the fact that the space-time affine connection is the usual of GR. Moreover,   the product of the metric and the Ricci scalar also turns out to be momentum independent
\begin{equation}
\begin{split}
&g_{\mu \nu}(x,k) R(x,k)\,=\,g_{\mu \nu}(x,k)g^{\rho \sigma}(x,k) R_{\rho \sigma}(x)\,=\,\\ &\left(1-\frac{\bar{k}_0^2-\vec{\bar{k}}^2}{4\Lambda^2}\right)^2 g^x_{\mu \nu}(x) \left(1-\frac{\bar{k}_0^2-\vec{\bar{k}}^2}{4\Lambda^2}\right)^{-2} g^{\rho \sigma}_x(x)  R_{\rho \sigma}(x)\,=\,g^x_{\mu \nu}(x) R^x(x)\,,
\end{split}
\end{equation}
due to the particular form of the metric~\eqref{eq:physical_metric} 
\begin{equation}
g_{\mu \nu}(x,k)\,=\,\left(1-\frac{\bar{k}_0^2-\vec{\bar{k}}^2}{4\Lambda^2}\right)^2 g^x_{\mu \nu}(x) \,,
\label{eq:metric_cotangent_definition}
\end{equation}
where $g^x_{\mu \nu}(x)$ and $ R^x(x)$ are the metric and scalar curvature in GR. This implies that the Einstein's equations in this scheme take the usual expression of GR. Summarizing, we have obtained a momentum basis for which the same energy-momentum tensor used in GR (momentum independent) can be used in this scheme, allowing us to generalize in a simple way how to consider curved spacetimes with this momentum deformation of the metric.

\subsection{About the choice of momentum basis}
\label{sec:choice_mb}
Here we discuss the invariance of our model about momentum changes of coordinates. As we have explained in Sec.~\ref{sec:construction_metric_ps}, in this framework the metric, and all tensors in general,  are invariant under diffeomorphisms. We can now see what happens if we apply a canonical transformation of the form
\be
k_\mu \,=\, h_\mu(k')\,,\quad x^\mu \,=\, x^{\prime \nu} j^\mu_\nu(k')\,,
\label{eq:primes}
\ee
 with
\be
j^\mu_\rho(k') \frac{\partial h_\nu(k')}{\partial k'_\rho} \,=\, \delta^\mu_\nu \,,
\ee
to the line element in the cotangent bundle~\eqref{eq:line_element_ps} when the metric does not depend on the space-time coordinates
\begin{equation}
\begin{split}
&g^\prime_{\mu\nu}(k^\prime) dx^{\prime\mu} dx^{\prime\nu}+g^{\prime \mu\nu}(k^\prime) d k^\prime_\mu d k^\prime_\nu\,=\,\frac{\partial x^{\prime\mu}}{\partial x^\rho} g^\prime_{\mu\nu}(k^\prime) \frac{\partial x^{\prime\nu}}{\partial x^\sigma} dx^{\rho} dx^{\sigma} \\
&+\frac{\partial k^\prime_\mu}{\partial k_\rho} g^{\prime \mu\nu}(k^\prime)\frac{\partial k^\prime_\nu}{\partial k_\sigma} d k_\rho d k_\sigma\,=\,g_{\mu\nu}(k) dx^{\mu} dx^{\nu}+g^{ \mu\nu}(k) d k_\mu d k_\nu\,.
\end{split}
\label{eq:line_element_ps_flat} 
\end{equation}
We can check that, for the flat space-time scenario, the metric line element is invariant. However, this cannot be done for the general case of a curved spacetime.  This means that our model presents a degeneracy for flat space-time cases in such a way that any coordinates in momentum space can be used. Nevertheless, for a curved space-time  this is no longer the case, and from the conservation of  Einstein's equations defined in~\cite{miron2001geometry} we find one and only one momentum basis which satisfies this requirement. Note that this degeneracy appears also when the momentum metric is flat, making that the usual variables of SR (in which the conservation law for momenta is the sum) are the only ones in which the Einstein's equations are conserved.

\section{Congruence of geodesics and Raychaudhuri's equation}
\label{sec:raychaudhuri}
We can now study the  Raychaudhuri's equation for a metric in the cotangent bundle~\cite{Relancio:2020zok,Relancio:2020rys}. For alleviate the notation, we shall not write explicitly the space-time and momentum dependence of the tensors. Along this section, we will follow the computations carried out in~\cite{Poisson:2009pwt}.  We start by computing the expansion of both timelike and null geodesics by its definition from the metric. The expansion of timelike geodesics is 
\begin{equation}
\theta \,=\,\frac{1}{\delta V}\frac{d }{d \tau}\delta V\,,
\label{eq:ge_volume}
\end{equation}
where $\delta V$ is the infinitesimal change of volume, while for null geodesics one has
\begin{equation}
\theta \,=\,\frac{1}{\delta S}\frac{d }{d \tau}\delta S\,,
\label{eq:ge_surface}
\end{equation}
where $\delta S$ is the infinitesimal change of surface. 
 
Now we consider a set of timelike geodesics labeled by $y^a\,(a=1,2,3)$ for each point in the set. This construction therefore defines a coordinate system $(\tau,y^a)$ in a neighborhood of the geodesic $\gamma$, and there exists a transformation between this system and the one originally in use: $x^\alpha=x^\alpha(\tau,y^a)$.  We define the vectors 
\begin{equation}
e^\alpha_a\,=\, \left(\frac{\partial x^\alpha}{\partial y^a}\right)_{\tau}\,.
\end{equation}  

Given the above definitions it is legitimate to ask (as usual) that the modified Lie derivative, Eq.~\eqref{eq:lie_vec}, for the velocity vector $u^\mu=d x^\mu/d\tau$ and $e^\alpha_a$ vanishes, 
\begin{equation}
\mathcal{L}_e u^\mu\,=\,e^\nu_a u^\mu_{\,;\nu}-u^\nu e^\mu_{a\,;\nu}-\frac{\partial u^\mu}{\partial k_\alpha} \frac{\partial e^\gamma_a}{\partial x^\alpha} k_\gamma -\frac{\partial u^\mu}{\partial k_\alpha}N_{\alpha \gamma}e^\gamma_a\,=\,0\,,
\label{eq:lie_vec2}
\end{equation}
we then find the following relation 
\begin{equation}
u^\mu e^\nu_{a\,;\mu}\,=\,e^\mu_a u^\nu_{\,;\nu}-\frac{\partial u^\mu}{\partial k_\alpha} \frac{\partial e^\gamma_a}{\partial x^\alpha} k_\gamma -\frac{\partial u^\mu}{\partial k_\alpha}N_{\alpha \gamma}e^\gamma_a\,.
\end{equation}

We can check that 
\begin{equation}
\begin{split}
\frac{d}{d \tau}\left(e^\nu_a u_\nu\right)\,&=\,\left(e^\nu_a u_\nu\right)_{;\mu}u^\mu\,=\,e^\nu_{a;\mu} u_\nu u^\mu+e^\nu_a u_{\nu;\mu}u^\mu  \\
\,&=\,\left(e^\mu_a u^\nu_{\,;\mu}-\frac{\partial u^\nu}{\partial k_\alpha} \frac{\partial e^\gamma_a}{\partial x^\alpha} k_\gamma -\frac{\partial u^\nu}{\partial k_\alpha}N_{\alpha \gamma}e^\gamma_a\right) u_\nu \\
\,&=\,\frac{1}{2}\left(e^\mu_a \left(u^\nu u_\nu\right)_{\,;\mu}-\frac{\partial \left(u^\nu u_\nu\right)}{\partial k_\alpha} \frac{\partial e^\gamma_a}{\partial x^\alpha} k_\gamma -\frac{\partial\left(u^\nu u_\nu\right)}{\partial k_\alpha}N_{\alpha \gamma}e^\gamma_a\right)\,=\,0\,,
\end{split}
\end{equation}
where we have used Eqs.~\eqref{cov_der_curve}-\eqref{eq:geodesic_cov_curve} and the fact that $u^\nu u_\nu$ is constant. Therefore, it is always possible to chose the parametrization of the geodesics in such a way that  $e^\nu_a u_\nu=0$, being $e^\nu_a$ orthogonal to the curves. 

We now introduce a three-tensor\footnote{A three-tensor is a tensor with respect to coordinate transformations $y^a \rightarrow y^{a'}$, but a scalar with respect to transformations $x^\alpha\rightarrow x^{\alpha'}$ if $y^a$ does not change.}  $h_{ab}$ defined by 
\begin{equation}
h_{ab}\,=\, g_{\alpha\beta}e^\alpha_a e^\beta_b\,.
\end{equation}  
This acts as a metric tensor on the infinitesimal volume we are considering: for displacements confined to the cross section (so that $d\tau=0$), $x^\alpha=x^\alpha(y^a)$ and 
\begin{equation}
ds^2\,=\,g_{\alpha\beta}dx^\alpha dx^\beta\,=\,g_{\alpha\beta}\frac{\partial x^\alpha}{\partial y^a}\frac{\partial x^\beta}{\partial y^b}dy^a dy^b\,=\,h_{ab}dy^a dy^b\,.
\end{equation}  
Thus, $h_{ab}$ is the three-dimensional metric on the congruence's cross sections. Because $\gamma$ is orthogonal to its cross sections $(u_\alpha e^\alpha_a=0)$, we have that $h_{ab}=h_{\alpha\beta}e^\alpha_a e^\beta_b$ on $\gamma$, where $h_{\alpha\beta}=g_{\alpha\beta}-u_\alpha u_\beta$ is the transverse metric. If we define $h^{ab}$ to be the inverse of $h_{ab}$, therefore one can easily  check that 
\begin{equation}
h^{\alpha\beta}\,=\,h^{ab}e^\alpha_a e^\beta_b\,
\end{equation}  
on $\gamma$.

The three-dimensional volume element on the cross sections is $\delta V=\sqrt{h}\,d^3y$, where $h=\det (h_{ab})$. Because the coordinates $y^a$ are comoving (since each geodesic moves with a constant value of its coordinates), $d^3y$ does not change as the cross section evolves. A change in $\delta V$ therefore comes entirely from a change in $\sqrt{h}$: 
 \begin{equation}
\theta \,=\,\frac{1}{\delta V}\frac{d }{d \tau}\delta V \,=\,\frac{1}{\sqrt{h}}\frac{d }{d \tau}\sqrt{h} \,=\,\frac{1}{2}h^{ab}\frac{d h_{ab} }{d \tau}\,.
\end{equation}

From the derivative of the three-metric we find
 \begin{equation}
 \begin{split}
\theta \,=\,&\frac{1}{2}h^{ab}\frac{d h_{ab} }{d \tau}\,=\,\frac{1}{2} e^a_\gamma e^b_\delta g^{\gamma\delta}g_{\alpha\beta}\left(e^\alpha_{a\,; \mu}u^\mu e^\beta_b+e^\beta_{b\,; \mu} u^\mu e^\alpha_a\right)\\
\,=\,&e^a_\alpha\left( u^\alpha_{\,;\mu}e^\mu_ a -\frac{\partial u^\alpha}{\partial k_\rho}N_{\gamma \rho}e^{\gamma}_a -\frac{\partial u^\alpha}{\partial k_\rho}\frac{\partial e^{\gamma}_a}{\partial x^\rho}k_\gamma  \right) \,=\,u^\mu_{\,;\mu}\,,
 \end{split}
\label{eq:ray_derivation}
\end{equation}
where we have use that $e_\mu^a$ does not depend on $k$ and $e_\alpha^a  u^\alpha=0$.

It is possible then to define, as in the GR case, the tensor
\begin{equation}
B_{\alpha\beta}\,=\,u_{\alpha;\beta}\,,
\label{eq:tensor_B}
\end{equation}
where $u^\mu$ is the tangent vector to the geodesics and  the new covariant derivative of Eq.~\eqref{eq:cov_dev_st} was used. Therefore, the tensor $B_{\alpha\beta}$ can be decomposed into trace, symmetric-trace free and antisymmetric parts
\begin{equation}
B_{\alpha\beta}\,=\,\frac{1}{3} \theta h_{\alpha\beta}+\sigma_{\alpha\beta}+\omega_{\alpha\beta}\,,
\end{equation}
being
\begin{equation}
\theta\,=\,B^\alpha _\alpha\,,\qquad\sigma_{\alpha\beta}\,=\,\frac{1}{2}\left(B_{\alpha\beta}+B_{\beta\alpha}\right)-\frac{1}{3} \theta h_{\alpha\beta}\,,\qquad \omega_{\alpha\beta}\,=\,\frac{1}{2}\left(B_{\alpha\beta}-B_{\beta\alpha}\right)\,.
\label{eq:theta1}
\end{equation}

The last step for obtaining the Raychaudhuri's equation is to compute the derivative of $\theta$ with respect to $\tau$. As we saw in Eq.~\eqref{eq:geodesic_cov_curve}, the following identity holds
\begin{equation}
\frac{d \theta}{d \tau}\,=\,\theta_{;\nu}u^\nu\,=\,u^\mu_{;\mu;\nu}u^\nu\,.
\end{equation}
Finally, using the commutator of two covariant derivative acting on a covariant vector~\eqref{commutator_cov_der2}, one finds
\begin{equation}
\frac{d \theta}{d \tau}\,=\, -B^{\alpha \beta}B_{\alpha \beta}-R_{\alpha\beta}u^\alpha u^\beta-  R_{\nu\beta\mu}(x,k)\frac{\partial u^\beta}{\partial k_\nu} u^\mu\,.
\end{equation}

One can check that the transverse metric satisfies
\begin{equation}
u^\alpha h_{\alpha\beta}\,=\,u^\beta h_{\alpha\beta}\,=\,0\,.
\end{equation}

We can express the  modified Raychaudhuri's equation using Eq.~\eqref{eq:theta1}, finding
\begin{equation}
\frac{d \theta}{d \tau}\,=\,-\frac{1}{3}\theta^2-\sigma^{\mu\nu}\sigma_{\mu\nu}+\omega^{\mu\nu}\omega_{\mu\nu}-R_{\mu\nu}u^\mu u^\nu- g^{\alpha \beta} R_{\nu\beta\mu}\frac{\partial u_\alpha}{\partial k_\nu} u^\mu\,.
\label{eq:r_timelike}
\end{equation}
This is the Hamiltonian version of the corresponding formula appearing in the Finslerian context~\cite{Stavrinos:2016xyg}.

For the null case we start by noticing that, for the momentum basis we are choosing~\eqref{eq:physical_metric}, a null vector for a generic curved spacetime satisfies
\begin{equation}
u_\alpha g^{\alpha\beta}(x,k)u_\beta\,=\,0\,\implies\,u_\alpha g_x^{\alpha\beta}(x)u_\beta\,\,=\, 0\,,
\end{equation}
where, as previously, $g_x^{\alpha\beta}(x)$ is the usual GR metric. This construction of a momentum independent null vector can be always done for the conformally flat metric~\eqref{eq:physical_metric} considered here. For the timelike case, as we will see in the next section, one finds nontrivial terms involving the mass of the particle probing spacetime.  For a generic momentum dependency of the metric, this simple choice cannot be used. 

Therefore we can introduce the usual auxiliary null vector $N_\alpha$, leading to the transverse metric~\cite{Poisson:2009pwt}
\begin{equation}
h_{\alpha\beta}\,=\, g_{\alpha\beta}-N_\alpha u_\beta-u_\alpha N_\beta\,,
\end{equation}
where
\begin{equation}
u^\beta h_{\alpha\beta}\,=\,N^\beta h_{\alpha\beta}\,=\,0\,,\qquad u^\alpha N_\alpha\,=\,1\,,\qquad N_\mu N^\mu\,=\,0\,.
\label{eq:n}
\end{equation}
In this case, one can define the purely traverse part of $B_{\alpha\beta}$ as   
\begin{equation}
\tilde{B}_{\alpha\beta}\,=\,h^\mu_{\alpha}h^\nu_{\beta}B_{\mu\nu}\,,
\end{equation}
whose decomposition is 
\begin{equation}
\tilde{B}_{\alpha\beta}\,=\,\frac{1}{2} \theta h_{\alpha\beta}+\sigma_{\alpha\beta}+\omega_{\alpha\beta}\,,
\end{equation}
where
\begin{equation}
\theta\,=\,\tilde{B}^\alpha _\alpha\,,\qquad\sigma_{\alpha\beta}\,=\,\frac{1}{2}\left(\tilde{B}_{\alpha\beta}+\tilde{B}_{\beta\alpha}\right)-\frac{1}{2} \theta h_{\alpha\beta}\,,\qquad \omega_{\alpha\beta}\,=\,\frac{1}{2}\left(\tilde{B}_{\alpha\beta}-\tilde{B}_{\beta\alpha}\right)\,.
\label{eq:theta2}
\end{equation}
The  Raychaudhuri's equation for null geodesics can be obtained as before from the commutator of two covariant derivatives
\begin{equation}
\frac{d \theta}{d \lambda}\,=\,-\frac{1}{2}\theta^2-\sigma^{\mu\nu}\sigma_{\mu\nu}+\omega^{\mu\nu}\omega_{\mu\nu}-R_{\mu\nu}u^\mu u^\nu\,.
\label{eq:r_null}
\end{equation}
The last term in Eq.~\eqref{eq:r_timelike} does not appear here due to the momentum independent expression for the null vector, leaving   the Raychaudhuri's equation unchanged with respect to the GR case.

\section{Friedmann-Lemaître-Robertson-Walker universe}
\label{sec:universe}
In this section, we study a flat expanding universe with the momentum dependence proposed in Sec.~\ref{sec:einstein}. The propose of this section is to study the congruence of geodesics and its relationship with the second Friedmann's equation~\cite{Relancio:2020rys}, a well-known fact in GR~\cite{Weinberg:1972kfs,Wald:1984rg}. But first, we need to obtain the vectors corresponding to  timelike and null geodesics. This can be done by using the equations of motion with a deformed Hamiltonian (see~\eqref{eq:H-eqs}) or through the geodesic equation~\eqref{eq:horizontal_geodesics_curve_definition}, obtaining the same result. 

We start from Eq.~\eqref{eq:metric_cotangent_definition} in order to construct our metric in the cotangent bundle, obtaining
\begin{equation}
\begin{split}
g_{00}(x,k)&\,=\,\left(1-\frac{k_0^2-\vec{k}^2/a^2(x^0)}{4\Lambda^2}\right)^2\,,\qquad g_{0i}(x,k)\,=\,0\,, \\ g_{ij}(x,k)&\,=\,\eta_{ij} a^2(x^0) \left(1-\frac{k_0^2-\vec{k}^2/a^2(x^0)}{4\Lambda^2}\right)^2\,.
\end{split}
\label{eq:RW_metric2}
\end{equation}

The Casimir can be obtained from the relation~\eqref{eq:casimir_definition_cst},
\begin{equation}
C(\bar{k})\,=\,4\Lambda^2 \arccoth^2\left(\frac{2 \Lambda}{\sqrt{k_0^2-\vec{k}^2/a^2(x^0)}}\right)\,.
\label{eq:casimir_RW}
\end{equation}
As expected, for $\Lambda$ going to infinity one recovers the usual expression of GR.

As discussed in Sec.~\ref{sec:einstein}, for the particular choice of momentum metric we use,  the space-time affine connection is the same one of GR~\cite{Weinberg:1972kfs}
 \begin{equation}
{H^0}_{ii}(x)\,=\,a(x^0)a^\prime(x^0)\,,\quad {H^i}_{0i}(x)\,=\,\frac{a^\prime(x^0)}{a(x^0)}\,,\quad {H^i}_{ij}(x)\,=\,0\,,\quad {H^i}_{jj}(x)\,=\,0\,,
\end{equation}
and also the nonlinear coefficients 
 \begin{equation}
N_{\mu\nu}(x,k)=\,k_{\lambda}{H^\lambda}_{\mu\nu}(x)\,.
\end{equation}
 
\subsection{Equations of motion from Hamilton equations}

From Casimir~\eqref{eq:casimir_RW} we can obtain the relationship between the components of the momentum for the massive case (in $1+1$ dimensions for simplicity)
\begin{equation}
k_0\,=\,\frac{1}{a(x^0)}\sqrt{k_1^2 \coth\left(\frac{m}{2\Lambda}\right) + 4\Lambda^2 a^2(x^0)}\tanh\left(\frac{m}{2\Lambda}\right)\,,
\label{eq:energy_massive}
\end{equation} 
and for the massless case
\begin{equation}
k_0\,=\,\frac{k_1}{a(x^0)}\,.
\label{eq:energy_massless}
\end{equation} 
As discussed above, for the massless scenario we obtain the same result of GR. This is due to the fact that the metric, and then also the deformed Casimir, is proportional to the mass of the particle probing the spacetime. Therefore, there is no modification at all for the massless case with respect to the computations of GR.

The velocity for massive particles can be obtained from the first equation of~\eqref{eq:H-eqs}, finding
\begin{equation}
\begin{split}
\dot{x}^0&\,=\,\cosh^2\left(\frac{m}{2\Lambda}\right)\sqrt{\frac{k_1^2}{4 \Lambda^2 a^2(x^0)} \coth^2\left(\frac{m}{2\Lambda}\right)+1}\,,\\ \dot{x}^1&\,=\,-\frac{k_1}{2 \Lambda a^2(x^0)}\cosh^2\left(\frac{m}{2\Lambda}\right)\coth\left(\frac{m}{2\Lambda}\right)\,,
\end{split}
\label{eq:massive_velocities}
\end{equation} 
where Eq.~\eqref{eq:energy_massive} has been used. For the massless case we get
\begin{equation}
 \frac{dx^0}{d\tau}\,=\,k_0\,=\,-\frac{k_1}{a(x^0)}\,,\qquad  \frac{dx^1}{d\tau}\,=\,-\frac{k_1}{ a^2(x^0)}\,,
\label{eq:massless_velocities}
\end{equation} 
where in this case Eq.~\eqref{eq:energy_massless} has been applied.

Now, from the the second equation of~\eqref{eq:H-eqs} we find 
\begin{equation}
\ \frac{dk_0}{d\tau}\,=\,\frac{d k_0}{d x^0}\dot{x}^0 \,=\, \left(k_0^2 \coth^2\left(\frac{m}{2\Lambda}\right)-4\Lambda^2 \right)\frac{a^{\prime}(x^0)}{4 \Lambda a(x^0)}\sinh\left(\frac{m}{2\Lambda}\right)\,, \qquad   \frac{dk_1}{d\tau}\,=\,0\,,
\label{eq:momenta_RW}
\end{equation}
for massive particles. It is easy to check that Eq.~\eqref{eq:energy_massive} can be recovered  from the first equation of Eq.~\eqref{eq:momenta_RW}.

For the massless case one finds
\begin{equation}
\frac{dk_0}{d\tau}\,=\, k_0^2\frac{a^{\prime}(x^0)}{a(x^0)}\,, \qquad  \dot{k}_1\,=\,0\,.
\label{eq:momenta_RW2}
\end{equation}
Again, Eq.~\eqref{eq:energy_massless} can be obtained by solving these equations.

\subsection{Equations of motion from the metric}

We can now obtain the same results from the geodesic equation Eq.~\eqref{eq:horizontal_geodesics}, whose components are
\begin{equation}
\frac{d^2x^0}{d\tau^2}+a(x^0)R'(x^0)\left(\frac{dx^1}{d\tau}\right)^2\,=\,0\,,\qquad \frac{d^2x^1}{d\tau^2}+2\frac{dx^1}{d\tau}\frac{R'(x^0)}{a(x^0)} \frac{dx^0}{d\tau}\,=\,0\,.
\label{eq:geo_RW}
\end{equation} 

For  massive particles, the line element for a horizontal curve is 
\begin{equation}
1\,=\,\frac{dx^\mu}{d\tau}g_{\mu\nu}\frac{dx^\nu}{d\tau}\,=\,\left(1-\frac{k_0^2-\vec{k}^2/a^2(x^0)}{4\Lambda^2}\right)^2\left(\left(\frac{dx^0}{d\tau}\right)^2-a^2(x^0)\left(\frac{dx^1}{d\tau}\right)^2\right)\,.
\label{eq:velocity_RW_massive}
\end{equation} 
Therefore, using~\eqref{eq:energy_massive} and the first equation of~\eqref{eq:geo_RW} one gets
\begin{equation}
\frac{d^2x^0}{d\tau^2}\,=\,\frac{dx^0 }{d\tau}\frac{d }{d x^0}\frac{dx^0 }{d\tau}\,=\,\frac{R'(x^0)}{a(x^0)}\cosh^4\left(\frac{m}{2\Lambda}-(u^0)^2\right)\,.
\end{equation}
The solution found in Eq.~\eqref{eq:massive_velocities} can be obtained by solving this last equation when  choosing $\tau$ to be the proper time. It is easy to see that the identification with    $\tau$ as the temporal coordinate cannot be done in this scheme since  one would have an imaginary solution for the velocity in  Eq.~\eqref{eq:velocity_RW_massive}. However, we can choose a parametrization in which $dx^0/d\tau$ does not depend on the momentum explicitly, finding
\begin{equation}
\frac{dx^0 }{d\tau}\,=\,\cosh^2\left(\frac{m}{2\Lambda}\right)\,,
\label{eq:massive_velocity_2}
\end{equation} 
which reduces to $dx^0/d\tau=1$  when the high-energy scale goes to infinity.

Moreover, from Eq.~\eqref{eq:horizontal_momenta} we get
\begin{equation}
\frac{dk_0}{d\tau}\,=\,k_1\frac{R'(x^0)}{a(x^0)}\frac{dx^1}{d\tau}\,,\qquad 
\frac{dk_1}{d\tau}\,=\,k_1\frac{ R'(x^0)}{a(x^0)}\frac{dx^0}{d\tau}+k_0 a(x^0)R'(x^0)\frac{dx^1}{d\tau}\,.
\end{equation} 
One can check that that these equations lead to the same solution found in Eq.~\eqref{eq:momenta_RW}.

For the case of photons, we have
\begin{equation}
0\,=\,\left(1-\frac{k_0^2-k_1^2/a^2(x^0)}{4\Lambda^2}\right)^2\left(\left(\frac{dx^0}{d\tau}\right)^2-a^2(x^0)\left(\frac{dx^1}{d\tau}\right)^2\right)\,,
\label{eq:velocity_RW}
\end{equation} 
and again, Eq.~\eqref{eq:massless_velocities} can be recovered from this expression. Moreover, Eq.~\eqref{eq:horizontal_momenta} can be also  deduced from Eq.~\eqref{eq:momenta_RW2} when  using Eq.~\eqref{eq:energy_massless}.

With the computations carried out in these two subsections we have checked explicitly that the same results are obtained from the Hamilton equations with a deformed Casimir, when it is identified with a function of the squared distance in momentum space, and  from the geodesic equation, with the space-time affine connection defined in~\cite{miron2001geometry}. This is consistent with the discussion of Sec.~\ref{sub:geo}.

\subsection{Congruence of geodesics and Raychaudhuri's equation}
Now we are able to apply the results of Sec.~\ref{sec:raychaudhuri} for obtaining  the congruence of geodesics and Raychaudhuri's equation for both timelike and null geodesics in a momentum dependent  FLRW spacetime. 

\subsubsection{Timelike geodesics}

Due to the fact that the volume  is proportional to the mass of the particle probing the spacetime, it disappears in Eq.~\eqref{eq:ge_volume}. We then find the term proportional to the mass of Eq.~\eqref{eq:massive_velocity_2}  multiplying the GR result 
\begin{equation}
\theta \,=\,\frac{1}{\delta V}\frac{dx^0}{d \tau}\frac{d \delta V }{d x^0}\,=\,\frac{3 a^\prime(x^0)}{2 a(x^0)} \cosh^2\left(\frac{m}{2\Lambda}\right)\,.
\label{eq:theta_massive}
\end{equation}
Here we have used the choice of the $\tau$ parameter of Eq.~\eqref{eq:massive_velocity_2}  in order to obtain the same GR result~\cite{Poisson:2009pwt} when $\Lambda$ goes to infinity. 

It is now easy to compute its derivative
\begin{equation}
\frac{d \theta}{d \tau} \,=\,\frac{\delta \theta}{\delta x^\mu}u^\mu\,=\,\, 3\frac{a^{\prime\prime}(x^0)a(x^0)-a^{\prime 2}(x^0)}{a^2(x^0)} \cosh^4\left(\frac{m}{2\Lambda}\right)\,.
\label{eq:theta_massive_der}
\end{equation}

The Raychaudhuri’s equation can be now obtained  by considering the tangent vector field 
\begin{equation}
u_\mu\,=\,\left(1-\frac{k_0^2-\vec{k}^2/a^2(x^0)}{4\Lambda^2},0,0,0\right)\,,
\end{equation} 
so $u^\mu u_\mu=1$. From Eq.~\eqref{eq:tensor_B} and the first equation of~\eqref{eq:theta1} we obtain
\begin{equation}
\theta\,=\,\frac{12 \Lambda^2 a(x^0)a^\prime(x^0)}{\vec{k}^2-\left(k_0^2-4\Lambda^2 \right)a^2(x^0)}\,,
\end{equation}
and using Eq.~\eqref{eq:energy_massive} we find the same result of Eq.~\eqref{eq:theta_massive}. Also, from Eq.~\eqref{eq:r_timelike} one obtains the same result of Eq.~\eqref{eq:theta_massive_der}. Therefore, our construction of the Raychaudhuri’s equation is consistent with the definition of the congruence of geodesics. It is important to notice that the modification with respect to the GR result is only   a multiplicative factor depending on the mass of the particle, which agrees with our previous discussion about the fact that the momentum dependency of the metric is through it.

\subsubsection{Null geodesics}

We can now see what happens for null geodesics. From Eq.~\eqref{eq:ge_surface} we find
\begin{equation}
\theta\,=\,2\frac{a^\prime(x^0)}{a^2(x^0)}\,,
\label{eq:theta_massless}
\end{equation}
and its derivative is
\begin{equation}
\frac{d \theta}{d \lambda}\,=\,2\frac{a^{\prime \prime}(x^0)a(x^0)-a^{\prime 2}(x^0)}{R^4(x^0)}\,.
\label{eq:theta_massless_der}
\end{equation}

The Raychaudhuri’s equation can be obtained by choosing the null vector (which, as we have explained previously, does not depend on the momentum)
\begin{equation}
u_0\,=\,\frac{1}{a(x^0)}\,\qquad u_1\,=\,1\,\qquad u_2\,=\, u_3\,=\,0\,.
\end{equation}
One can check that the geodesic equation holds
\begin{equation}
u_{\mu;\lambda}\,u^\lambda\,=\,\left(\frac{\delta u_\mu}{\delta x^\lambda}-{H^\nu}_{\mu\lambda}u_\nu\right)\,u^\lambda\,=\,0\,,
\end{equation}
which is consistent with Eq.~\eqref{eq:geodesic_cov_curve}. As an auxiliary null vector we can take
\begin{equation}
N_0\,=\, -N_1\,=\,\frac{\left(\vec{k}^2-\left(k_0^2-4\Lambda^2\right)a^2(x^0)\right)^2}{32 \Lambda^4 a^2(x^0)}\,\qquad N_2\,=\, N_3\,=\,0\,,
\end{equation}
in such a way that Eq.~\eqref{eq:n} holds.  As expected, from Eq.~\eqref{eq:energy_massless} we find the same result of Eq.~\eqref{eq:theta_massless} and also, from Eq.~\eqref{eq:r_null} one obtains the same result of Eq.~\eqref{eq:theta_massless_der}. We find that there is no modification for the massless case, which is consistent with how the metric depends on the momentum. 

\subsection{Friedmann's equations}
In this subsection we check that the second Friedmann's equation can be recovered from the  Raychaudhuri's equations we have obtained previously.  The energy-momentum tensor for a FLRW universe is a perfect fluid
\begin{equation}
T_{\mu \nu}\,=\,u_\mu u_\nu \left(\rho(x^0)+P(x^0)\right)-P(x^0) g_{\mu \nu}(x)\,\qquad \text{with}\qquad u_\mu\,=\,\left(1,0,0,0\right)\,,
\label{eq:em_tensor_pf}
\end{equation}
where $u_\mu$ the fluid's four velocity, and $\rho$ and $P$ the energy density and pressure of the fluid, respectively.  As  discussed in Sec.~\ref{sec:einstein}, our choice of momentum metric~\eqref{eq:physical_metric} leads to same Ricci tensor of GR, and then, the space-time scalar of curvature obtained from Eq.~\eqref{eq:ricci_scalar} is
\begin{equation}
R\,=\,\frac{96 \Lambda^4 a^2(x^0)\left(a^{\prime 2}(x^0)+a(x^0)a^{\prime \prime}(x^0)\right)}{\left(a^2(x^0)\left(k_0 ^2-4\Lambda^2\right)-\vec{k}^2\right)^2} \,.
\end{equation}
Also, as explained in Sec.~\ref{sec:einstein}, due to the conformal factor of the momentum metric of Eq.\eqref{eq:physical_metric}, there is a cancellation of momentum factors in the Einstein's tensor leading to the same tensor one finds in GR. Therefore, there is no modification of the Einstein's equations, and then the same Friedmann's equations of GR (for both massive and massless particles) are valid in this scheme,
\begin{equation}
 a^{\prime 2}(x^0)\,=\,\frac{8 \pi}{3}a^2(x^0)\rho(x^0)\,,\qquad 2 a^{\prime \prime}(x^0)a(x^0)+ a^{\prime 2}(x^0)\,=\,8 \pi P(x^0)\,.
\label{eq:Friedmann_eqs}
\end{equation}

Since for the massless case there is no modification of the Raychaudhuri's equation, we focus in the massive scenario. By multiplying the Einstein's equations~\eqref{eq:einstein_eqs} by the inverse of the metric we find the same relationship as in GR (apart from the momentum dependence) between the scalar of curvature of spacetime and the scalar of the energy-momentum tensor
 \begin{equation}
R-2R\,=\, 8\pi T\implies\,R\,=\,- 8 \pi T\,.
\label{eq:r-t}
\end{equation}
Therefore,  using Eqs.~\eqref{eq:einstein_eqs}-\eqref{eq:r-t}, we can rewrite  the Raychaudhuri's equation for massive particles~\eqref{eq:r_timelike}  as 
 \begin{equation}
R_{\mu \nu} u^\mu u^\nu\,=\, \left(8 \pi T_{\mu \nu}+\frac{1}{2}g_{\mu \nu} R\right)u^\mu u^\nu\,=\,8 \pi \left( T_{\mu \nu}u^\mu u^\nu -\frac{1}{2}T\right)\,.
\end{equation}

We can now set the following relation between the congruence of geodesics and  the energy-momentum tensor
\begin{equation}
\begin{split}
8 \pi \left( T_{\mu \nu}u^\mu u^\nu -\frac{1}{2}T\right)\,=\,&\frac{d \theta}{d \tau}+\frac{1}{3}\theta^2+\sigma^{\mu\nu}\sigma_{\mu\nu}-\omega^{\mu\nu}\omega_{\mu\nu}+ g^{\alpha \beta} R_{\nu\beta\mu}\frac{\partial u_\alpha}{\partial k_\nu} u^\mu\\
\,=\,&3 \cosh^4\left(\frac{m}{2\Lambda}\right)\frac{a^{\prime\prime}(x^0)}{a(x^0)} \,.
\end{split}
\label{eq:friedmann12}
\end{equation}
Finally, taking into account the explicit form of the energy-momentum tensor~\eqref{eq:em_tensor_pf} one finds
\begin{equation}
8 \pi \left( T_{\mu \nu}u^\mu u^\nu -\frac{1}{2}T\right)\,=\, 4\cosh^4\left(\frac{m}{2\Lambda}\right)\left(3P(x^0)+\rho(x^0)\right)\,,
\end{equation}
which substituted in Eq.~\eqref{eq:friedmann12} leads to the same expression one obtains by combining the first and second Friedmann's equations. 

With this we have checked the fact that, as in GR, the second Friedmann's equation can be recovered from the Einstein's equations and from the Raychaudhuri’s equation.

\section{Conclusions}
\label{sec:conclusions}

In this review we have discussed how a relativistic deformed kinematics can be obtained from a curved momentum space, and how to extend this notion to curved spacetimes, leading to a geometry in the cotangent bundle. 

We have arrived to the conclusion that only a maximally symmetric momentum space could lead to a relativistic deformed kinematics when one identifies the composition law and the Lorentz transformations as the isometries of the metric~\cite{Carmona:2019fwf}. Since one wants 4 translations and 6 Lorentz generators, the momentum space must have 10 isometries, leaving only place for a maximally symmetric space. We have found that the most common examples of deformed kinematics appearing in the literature ($\kappa$-Poincaré, Snyder and hybrid models) can be reproduced and understood from this geometrical perspective by choosing properly the algebra of the generators of translations.

We have also investigated how a curvature of spacetime could modify the kinematics~\cite{Relancio:2020zok}. In order to do so, we have studied, from a geometrical point of view, how to consider simultaneously a curvature in spacetime and in momentum space. This can be done in the so-called cotangent bundle geometry, taking into account a nontrivial geometry for all the phase space.

Moreover, we have advanced in  several formal aspects about the consistent generalization of the description of a curved spacetime with the local group symmetry of a relativistic deformed kinematics. We have proved that by imposing the same description of the free particles motion  from both a deformed Hamiltonian (which is a function of the squared distance in momentum space) and the geodesic motion given by a geometry in the cotangent bundle,  leads to a nontrivial relation between the geometrical ingredients.  We have also studied the modification of the Lie derivative and the Raychaudhuri's equation. 

By imposing the conservation of the Einstein's tensor, so the energy-momentum tensor will be also a conserved quantity, we find a preferred system of momentum coordinates, which implies a particular basis of the deformed kinematics~\cite{Relancio:2020rys}. Therefore, accepting this geometrical setup for describing relativistic deformed kinematics, we find that this framework provides the breakdown of degeneracy of flat spacetime that appears in DSR, leading to a ``physical'' basis in which the laws of nature must be described. Whilst this basis is obtained when considering a curvature on spacetime (i.e., there is a nontrivial dependence in the space-time coordinates in the metric),  this basis should be also used for studying phenomenology in DSR context for flat spacetime, so having a smooth limit towards GR.

This particular momentum metric is conformally flat, with a prefactor which depends on the squared four-momentum divided by the scale of deformation~\cite{Relancio:2020rys}. Therefore, the dispersion relation, constructed as a function of the squared distance in momentum space, depends solely on the four-momentum squared. This leads to a modification depending on the mass of the particle probing spacetime, so for massless particles there would not be any modification of the trajectories. This could not be the case for massive particles, where a function of the squared mass divided by the high-energy scale arises. 

In this coordinates we revised the Friedmann-Lemaître-Robertson-Walker universe checking that the trajectories of massive and massless particles in both Hamiltonian and geometrical frameworks lead to the same results~\cite{Relancio:2020rys}. This corroborates our derivations presented along the paper. Also, we have constructed the modified Raychaudhuri's equation following the method carried out in GR. Moreover, we find that the second Friedmann's equation is compatible with our definition of the Einstein's and  Raychaudhuri's  equations, which is an important consistency check.  Since in this preferred basis the  Einstein's equations are not modified, the generalizations of any GR metric to this framework can be easily developed. 

Moreover, while in the one-particle sector could be difficult to observe a deviation of SR and GR (due to the covariant form of the metric, precluding a time delay of photons), there could be some interesting processes involving the collision of several particles. While in the SR context this has been started to be studied~\cite{Albalate:2018kcf,Carmona:2020whi,Carmona:2021lxr}, this is not the case for the curved spacetime scenario. The deformation of the kinematics in curved spacetimes could have phenomenological consequences for astrophysical processes, like collisions of particles near black holes.

In the end, we have understood how to consider a deformed metric in the cotangent bundle for the one-particle system for both flat and curves spacetimes. A geometrical structure for a multi-particle scheme was considered in~\cite{Relancio:2021ahm} only for flat spacetimes, being an open question its generalization to the curved case.
\section*{Acknowledgments}

The author acknowledges support from the INFN Iniziativa Specifica GeoSymQFT. The author would also like to thank support from the COST Action CA18108.

\end{document}